\RequirePackage[l2tabu,orthodox]{nag}
\documentclass
[11pt,letterpaper]
{article}

\usepackage[dvipsnames]{xcolor}

\usepackage[notes=false,later=false,camera=false]{dtrt}
\usepackage[utf8]{inputenc}
\usepackage{etex}
\usepackage{ stmaryrd }
\usepackage{xspace,enumerate}
\usepackage[T1]{fontenc}
\usepackage[full]{textcomp}
\usepackage[american]{babel}
\usepackage{mathtools}

\usepackage{amsthm}
\usepackage{empheq}

   \usepackage{hyperref}
   \hypersetup{
colorlinks=true,
urlcolor=Cerulean,
linkcolor=RoyalBlue,
citecolor=OliveGreen,
linktocpage=true,
}
\usepackage[capitalise,nameinlink]{cleveref}
\crefname{lemma}{Lemma}{Lemmas}
\crefname{fact}{Fact}{Facts}
\crefname{theorem}{Theorem}{Theorems}
\crefname{mtheorem}{Theorem}{Theorems}
\crefname{corollary}{Corollary}{Corollaries}
\crefname{claim}{Claim}{Claims}
\crefname{example}{Example}{Examples}
\crefname{algorithm}{Algorithm}{Algorithms}
\crefname{problem}{Problem}{Problems}
\crefname{definition}{Definition}{Definitions}
\crefname{question}{Question}{Questions}
\usepackage{paralist}
\usepackage{turnstile}
\usepackage{mdframed}
\usepackage{tikz}
\usepackage{caption}
\DeclareCaptionType{Algorithm}
\usepackage{newfloat}
\newtheorem{theorem}{Theorem}[section]
\newtheorem{mtheorem}{Theorem}
\newtheorem*{theorem*}{Theorem}

\newtheorem{proposition}[theorem]{Proposition}
\newtheorem*{proposition*}{Proposition}
\newtheorem{lemma}[theorem]{Lemma}
\newtheorem*{lemma*}{Lemma}
\newtheorem{corollary}[theorem]{Corollary}
\newtheorem*{conjecture*}{Conjecture}
\newtheorem{fact}[theorem]{Fact}
\newtheorem*{fact*}{Fact}

\newtheorem*{hypothesis*}{Hypothesis}

\theoremstyle{definition}
\newtheorem{definition}[theorem]{Definition}
\newtheorem*{definition*}{Definition}

\newtheorem{question}[theorem]{Question}

\newtheorem{algorithm}[theorem]{Algorithm}

\theoremstyle{remark}
\newtheorem{claim}[theorem]{Claim}
\newtheorem*{claim*}{Claim}
\newtheorem{remark}[theorem]{Remark}
\newtheorem*{remark*}{Remark}
\newtheorem{observation}[theorem]{Observation}
\newtheorem*{observation*}{Observation}

\usepackage[
letterpaper,
top=1.2in,
bottom=1.2in,
left=1in,
right=1in]{geometry}
\usepackage{newpxtext} 
\usepackage{textcomp} 
\usepackage[varg,bigdelims]{newpxmath}
\usepackage[scr=rsfso]{mathalfa}
\usepackage{bm} 
\let\mathbb\varmathbb
\usepackage{microtype}
\definecolor{petergreen}{rgb}{0, 0.75, 0}
\usepackage{multirow}
\usepackage{thm-restate}

\allowdisplaybreaks
\newcommand{\FormatAuthor}[3]{
\begin{tabular}{c}
#1 \\ {\small\texttt{#2}} \\ {\small #3}
\end{tabular}
}

\newcommand{\N}{{\mathbb N}}

\newcommand{\abs}[1]{\lvert #1 \rvert}
\newcommand{\floor}[1]{\lfloor #1 \rfloor}

\newcommand{\eps}{\varepsilon}
\renewcommand{\epsilon}{\eps}
\newcommand{\defeq}{:=}
\newcommand{\F}{{\mathbb F}}

\DeclareMathOperator{\E}{\mathbb{E}}
\newcommand{\1}{\mathbf{1}}
\newcommand{\ip}[1]{\langle #1 \rangle}

\newcommand{\Bits}{\{0,1\}}

\newcommand{\im}{\operatorname{im}}

\newcommand{\cH}{\mathcal H}

\newcommand{\poly}{\operatorname{poly}}

\newcommand{\mper}{\,.}
\newcommand{\mcom}{\,,}

\newcommand{\cD}{\mathcal D}

\newcommand{\cQ}{\mathcal Q}

\newcommand{\cV}{\mathcal{V}}

\newcommand{\Dec}{\operatorname{Dec}}

\newcommand{\Code}{C}
\newcommand{\vspan}{\operatorname{span}}
\newcommand{\supp}{\operatorname{supp}}
\newcommand{\Had}{\operatorname{Had}}
\newcommand{\LIN}{\mathrm{LIN}}
\newcommand{\msg}{b}
\newcommand\given[1][]{\:#1\vert\:}
\newcommand{\cW}{\mathcal{W}}
\newcommand{\cU}{\mathcal{U}}
\newcommand{\cP}{\mathcal{P}}
\newcommand{\cL}{\mathcal{L}}

\begin{document}


\title{\texorpdfstring{\vspace{-1cm}Relaxed vs.\ Full Local Decodability with Few Queries:\\
Equivalence and Separations for Linear Codes}{Relaxed vs.\ Full Local Decodability with Few Queries:\\
Equivalence and Separations for Linear Codes}}

\author{
\begin{tabular}[h!]{cc}
     \FormatAuthor{Elena Grigorescu\thanks{Supported in part by NSF Grant CCF-2228814 while at Purdue University.}}{elena-g@uwaterloo.ca}{University of Waterloo}
     & \FormatAuthor{Vinayak M. Kumar\thanks{Supported by NSF Grant CCF-2312573, a Simons Investigator Award (\#409864, David Zuckerman), a Jane Street Fellowship, and a UT Austin Dean's Prestigious Fellowship Supplement.}}{vmkumar@cs.utexas.edu}{University of Texas at Austin} \\ 
      & \\
     \FormatAuthor{Peter Manohar\thanks{Supported by NSF Grant DMS-2424441.}}{pmanohar@ias.edu}{The Institute for Advanced Study}
     & \FormatAuthor{Geoffrey Mon\thanks{Supported by NSF Grant CCF-2312573, an NSF Graduate Research Fellowship (DGE-2137420), and a UT Austin Dean’s Prestigious Fellowship Supplement.}}{gmon@cs.utexas.edu}{University of Texas at Austin}
\end{tabular}
} %
\date{\today}


\maketitle

\vspace{-0.3cm}
\begin{abstract}
A locally decodable code (LDC) $\Code \colon \Bits^k \to \Bits^n$ is an error-correcting code that allows one to recover any bit of the original message with good probability while only reading a small number of bits from a corrupted codeword. A relaxed locally decodable code (RLDC) is a weaker notion where the decoder is additionally allowed to abort and output a special symbol $\bot$ if it detects an error. For a large constant number of queries $q$, there is a large gap between the blocklength $n$ of the best $q$-query LDC and the best $q$-query RLDC. Existing constructions of RLDCs achieve polynomial length $n = k^{1 + O(1/q)}$~\cite{BenSassonGHSV04,GurRR18,ChiesaGS20,AsadiS21,Goldreich24}, while the best-known $q$-LDCs only achieve subexponential length $n = 2^{k^{o(1)}}$~\cite{Yekhanin08,Efremenko09}. On the other hand, for $q = 2$, it is known that RLDCs and LDCs are equivalent~\cite{BlockBCGLZZ23}. We thus ask the question: what is the smallest $q$ such that there exists a $q$-RLDC that is not a $q$-LDC?

In this work, we show that any linear $3$-query RLDC is in fact a $3$-LDC, i.e., linear RLDCs and LDCs are equivalent at $3$ queries. More generally, we show for any constant $q$, there is a soundness error threshold $s(q)$ such that any linear $q$-RLDC with soundness error below this threshold must be a $q$-LDC. This implies that linear RLDCs cannot have ``strong soundness'' --- a stricter condition satisfied by linear LDCs that says the soundness error is proportional to the fraction of errors in the corrupted codeword --- unless they are simply LDCs.

In addition, we give simple constructions of linear $15$-query RLDCs that are not $q$-LDCs for any constant $q$, showing that for $q = 15$, linear RLDCs and LDCs are not equivalent.

We also prove nearly identical results for locally correctable codes and their corresponding relaxed counterpart.
\end{abstract}

\thispagestyle{empty}
\clearpage
 \microtypesetup{protrusion=false}
  \tableofcontents{}
  \microtypesetup{protrusion=true}

\thispagestyle{empty}
\clearpage

\pagestyle{plain}
\setcounter{page}{1}
\section{Introduction}
\label{sec:intro}
A binary locally decodable code (LDC) $\Code \colon \Bits^k \to \Bits^n$ is an error-correcting code that admits a local decoding algorithm with the following guarantee: when given access to a corrupted codeword $y$ obtained by corrupting $x = \Code(b)$ in a constant fraction of coordinates, the local decoder is able to recover any bit $b_i$ of the chosen message $b$ with good probability while only reading a small (say, constant) number of bits from $y$.
Although they were first formally defined in~\cite{KatzT00}, locally decodable codes were implicitly used in the proof of the PCP theorem~\cite{AroraLMSS98,AroraS98}, and have since found numerous applications to, e.g., private information retrieval,  hardness amplification,
probabilistically checkable proofs, self-correction,  fault-tolerant circuits and data structures (e.g., \cite{BabaiFLS91,LundFKN92,BlumLR93,BlumK95,ChorKGS98,IshaiK99,BarkolIW10,ChenGW13,AndoniLRW17}).
We refer the reader to the surveys of~\cite{Trevisan04,Dvir12,Yekhanin12} for more details.

The central question in the study of LDCs is to understand the length $n$ of the best locally decodable code that can tolerate a small constant fraction of errors as a function of $k$, the length of the message, and $q$, the number of queries of the decoder. Following~\cite{KatzT00}, there has been a long line of work on both constructing and proving lower bounds for locally decodable codes, with a particular focus on the constant query regime when $q = O(1)$.

A simple observation (see \cite[Section 3.2]{KatzT00}) shows that it is not possible to construct $1$-query locally decodable codes. For $q = 2$, the Hadamard code gives a $2$-LDC of blocklength $n = 2^k$, and this is optimal up to constant factors in the exponent: the works of~\cite{KerenidisW04,GoldreichKST06} show that $n \geq 2^{\Omega(k)}$ for any $2$-LDC. More generally, the Reed--Muller code --- a generalization of the Hadamard code (which are evaluations of linear functions) to polynomials of larger degree --- gives a construction of a $q$-LDC of length $n = 2^{O(k^{1/(q-1)})}$. However, for $q \geq 3$, this can be improved: the matching vector codes of~\cite{Yekhanin08,Efremenko09,DvirGY11} give constructions of $q$-LDCs of subexponential, but still superpolynomial, blocklength $n = 2^{k^{o(1)}}$ for any constant $q \geq 3$.\footnote{More precisely, the length of these codes is $n = 2^{2^{(\log k)^{\eps(q)}}}$ for some constant $\eps(q) \approx 1/\log_2 q < 1$ that depends on $q$.} But, unlike the case of $q = 2$, we are far from understanding whether these codes are optimal or not: our best lower bound is a polynomial lower bound of $n\geq \tilde{\Omega}(k^{q/(q-2)})$ for any $q \geq 3$~\cite{KerenidisW04,AlrabiahGKM23,HsiehKMMS24,BasuHKL25,JanzerM25}.

This gap between subexponential constructions and polynomial lower bounds for LDCs has led to the definition of a class of codes with weaker local decoding properties called relaxed locally decodable codes (RLDCs), for which much better constructions are known. Introduced in~\cite{BenSassonGHSV04}, a relaxed LDC is no longer required to output the correct message bit $b_i$ with good probability, and instead may output a special symbol $\bot$ to signify that the decoder has detected an error.\footnote{To prevent the trivial decoder that always outputs $\bot$ from satisfying the definition, one requires that the decoder outputs the correct bit $b_i$ with good probability when given access to an uncorrupted codeword $x = C(b)$.} Unlike standard locally decodable codes, for large enough constant $q$ one can construct $q$-query RLDCs with a polynomial blocklength of $n = k^{1 + O(1/q)}$~\cite{BenSassonGHSV04,GurRR18,ChiesaGS20,AsadiS21,Goldreich24}, and the works of~\cite{GurL20,DallAgnolGL21,Goldreich23} prove a near-matching lower bound of $n \geq k^{1 + \Omega(1/q^2)}$.
However, somewhat curiously, for the specific case of $q = 2$ the work of~\cite{BlockBCGLZZ23} proves an exponential lower bound of $n \geq 2^{\Omega(k)}$. That is, up to constant factors in the exponent, the best $2$-RLDC is a $2$-LDC, namely the Hadamard code.

To summarize: for RLDCs, there exists a large constant $q$ such that there are $q$-RLDCs of length $n = \poly(k)$, whereas the best $2$-RLDC has length $n = 2^{\Omega(k)}$. The work of \cite{BlockBCGLZZ23} thus raises the following interesting question (mentioned explicitly in \cite[Section 2]{BlockBCGLZZ23}):
\begin{question}[\cite{BlockBCGLZZ23}]
\label{ques:rldcphase}
What is the threshold $q$ where the optimal blocklength of a $q$-RLDC ``transitions'' from superpolynomial in $k$ to polynomial in $k$?
\end{question}
In this work, we investigate \cref{ques:rldcphase}. However, there is an obvious barrier to answering \cref{ques:rldcphase}, coming from our lack of understanding of the analogous question for LDCs. For example, if it is the case that the ``transition threshold'' is above $q = 3$, then to prove this one would need to prove that there are no $3$-RLDCs of polynomial length. In particular, this would also imply that there are no $3$-LDCs (unrelaxed) of polynomial length as well. However, the best $3$-LDC lower bound is only $n \geq \tilde{\Omega}(k^3)$~\cite{AlrabiahGKM23}, and improving this (or showing that it is tight) is a well-studied and challenging open question.

To avoid such issues, we instead reinterpret the result of \cite{BlockBCGLZZ23} as follows. Not only do they prove that any $2$-RLDC has $n \geq 2^{\Omega(k)}$, they in fact prove this result by showing that any $2$-RLDC \emph{is} a $2$-LDC, and then they apply the exponential lower bound of~\cite{KerenidisW04,GoldreichKST06} for $2$-LDCs. Thus, for $q = 2$, RLDCs and LDCs are equivalent. On the other hand, for large constant $q$ there is a large gap between the best constructions of $q$-RLDCs (which have length $n = k^{1 + O(1/q)}$~\cite{BenSassonGHSV04,AsadiS21,Goldreich24}) and $q$-LDCs (which have length $n = 2^{k^{o(1)}}$~\cite{Yekhanin08,Efremenko09,DvirGY11}), giving evidence that for a large enough constant $q$, RLDCs and LDCs are not equivalent. We thus pose the following question:
\begin{question}
\label{ques:rldcequiv}
What is the smallest $r$ where \begin{inparaenum}[(1)] \item for every $q < r$, every $q$-RLDC is a $q$-LDC, and \item there is an $r$-RLDC that is \emph{not} an $r$-LDC\end{inparaenum}?
\end{question}
Our main results show that the threshold $r$ in \cref{ques:rldcequiv} is between $4$ and $15$ for the case of linear codes. In fact, we show this for a ``stronger'' version of \cref{ques:rldcequiv} where the $r$-RLDC in Item (2) is not only not an $r$-LDC, but also not a $t$-LDC for any constant $t$.

\paragraph{Locally correctable and relaxed locally correctable codes.} The above discussion is for locally decodable codes and their relaxed variant. One may ask the same questions for locally correctable codes (LCCs) and their relaxed variant, as LCCs are a closely related notion to LDCs that are defined in a near-identical way. The difference between an LCC and an LDC is that a locally correctable code requires that the decoder is able to self-correct bits of the \emph{codeword}, whereas LDCs only need to correct bits of the \emph{message}. One can show that any LCC is in fact an LDC as well, so LCCs are a stronger notion.\footnote{This is fairly straightforward to show for linear codes, as without loss of generality, by changing bases one can make any linear code systematic, i.e., the first $k$ bits of any codeword $x = \Code(b)$ is simply the message $b$. For nonlinear codes, this can also be done with more effort, see~\cite[Appendix A]{BhattacharyyaGT17}.} And, similarly to LDCs, one can define a relaxed notion of LCCs where the decoder is allowed to output a special error symbol $\bot$.

For LCCs, the state-of-the-art constructions and lower bounds are quite different compared to LDCs. For $q = 2$, these notions are equivalent, and as shown by~\cite{KerenidisW04,GoldreichKST06}, the Hadamard code with $n = 2^k$ is also an optimal $2$-LCC. For $q \geq 3$, however, the best construction of $q$-LCCs remains the folklore construction from Reed--Muller codes, which achieves a length of $n = 2^{O(k^{1/(q-1)})}$. This is unlike the case of LDCs, where the constructions of subexponential length coming from matching vector codes are much better than Reed--Muller codes when $q \geq 3$.

We also have stronger lower bounds for LCCs as compared to LDCs. Namely, for $q = 3$, the work of~\cite{KothariM23} and follow-up works of~\cite{AlrabiahG24,Yankovitz24,KothariM24} prove exponential lower bounds for $3$-LCCs: the current best results are $n \geq 2^{\Omega(\sqrt{k/\log k})}$ for linear codes~\cite{AlrabiahG24} and $n \geq 2^{\Omega(k^{1/5})}$ for nonlinear codes~\cite{KothariM24}. For comparison, recall that the best $3$-LDC lower bound remains $n \geq \tilde{\Omega}(k^3)$. Furthermore, for any odd constant $q \geq 5$, one can show a lower bound of $n \geq \tilde{\Omega}(k^{(q-1)/(q-3)})$ for linear $q$-LCCs~\cite{AlrabiahG24}, which is better than the best $q$-LDC lower bound by a small polynomial factor in $k$.

For relaxed LCCs (RLCCs), the current best results are identical to RLDCs. Namely, for large enough constant $q$, the work of~\cite{AsadiS21} constructs $q$-query RLCCs of length\footnote{The works of~\cite{BenSassonGHSV04,Goldreich24} construct $q$-RLDCs with the same parameters, but their codes are not RLCCs.} $n = k^{1 + O(1/q)}$ and the lower bounds of~\cite{GurL20,DallAgnolGL21,Goldreich23} again prove a near-matching lower bound of $n \geq k^{1 + \Omega(1/q^2)}$. And, for $q = 2$, \cite{BlockBCGLZZ23} proves an exponential lower bound for $2$-RLCCs of $n \geq 2^{\Omega(k)}$.\footnote{Unlike the case of RLDCs/LDCs, the work of~\cite{BlockBCGLZZ23} does not prove a $2$-RLCC to $2$-LCC reduction. However, since any $2$-RLCC is a $2$-RLDC, the exponential lower bound still applies.}

We can thus also ask \cref{ques:rldcequiv} for LCCs and RLCCs.
\begin{question}[\cref{ques:rldcequiv} for LCCs]
\label{ques:rlccequiv}
What is the smallest $r$ where \begin{inparaenum}[(1)] \item for every $q < r$, every $q$-RLCC is a $q$-LCC, and \item there is an  $r$-RLCC that \emph{not} an $r$-LCC\end{inparaenum}?
\end{question}
We also prove results for LCCs/RLCCs similar to the case of LDCs/RLDCs. In particular, we show that the threshold $r$ in \cref{ques:rlccequiv} is between $4$ and $41$ for the case of linear codes.

\subsection{Our results}
Before we state our results, let us formally define LDCs and RLDCs (LCCs and RLCCs are defined analogously, see \cref{def:lcc,def:rlcc}). For two strings $x,y \in \Bits^n$, we let $\Delta(x,y)$ denote the Hamming distance between $x$ and $y$, i.e., the number of indices $j \in [n]$ where $x_j \ne y_j$.

\begin{definition}[Binary locally decodable codes; 
see \cref{def:ldc}]
  \label{def:binary-ldc}
A code $\Code \colon \Bits^k \to \Bits^n$ is a $(q, \delta, c, s)$-LDC if there exists a randomized decoding algorithm $\Dec(\cdot)$ with the following properties. The algorithm $\Dec(\cdot)$ is given oracle access to a string $y \in \Bits^n$, takes an index $i \in [k]$ as input, and outputs a bit in $\Bits$ with the following guarantees:
\begin{enumerate}[(1)]
\item ($q$-queries) for any $y$ and $i$, the algorithm $\Dec^{y}(i)$ reads at most $q$ indices of $y$,
\item ($c$-completeness) for all $b \in \Bits^k$ and $i \in [k]$, $\Pr[\Dec^{\Code(b)}(i) = b_i] \geq c$, and
\item ($(\delta, s)$-soundness error) for all $b \in \Bits^k$, $i \in [k]$, and all $y \in \Bits^n$ with $\Delta(y, \Code(b)) \leq \delta n$, $\Pr[\Dec^{y}(i) \ne b_i] \leq s$.
\end{enumerate}
\end{definition}

\begin{definition}[Binary relaxed locally decodable codes; see \cref{def:rldc}]
  \label{def:binary-rldc}
A code $\Code \colon \Bits^k \to \Bits^n$ is  a $(q, \delta, c, s)$-RLDC if there exists a randomized decoding algorithm $\Dec(\cdot)$ with the following properties. The algorithm $\Dec(\cdot)$ is given oracle access to a string $y \in \Bits^n$, takes an index $i \in [k]$ as input, and outputs either a bit in $\Bits$ or a special symbol $\bot$ with the following guarantees:
\begin{enumerate}[(1)]
\item ($q$-queries) for any $y$ and $i$, the algorithm $\Dec^{y}(i)$ reads at most $q$ indices of $y$,
\item ($c$-completeness) for all $b \in \Bits^k$ and $i \in [k]$, $\Pr[\Dec^{\Code(b)}(i) = b_i] \geq c$, and
\item ($(\delta, s)$-relaxed soundness error) for all $b \in \Bits^k$, $i \in [k]$, and all $y \in \Bits^n$ with $\Delta(y, \Code(b)) \leq \delta n$, $\Pr[\Dec^{y}(i) \notin \{b_i, \bot\}] \leq s$.
\end{enumerate}
\end{definition}
\noindent The standard definition of an RLDC sets parameters as follows: $\delta = O(1)$ (constant fraction of errors), $c = 1$ (perfect completeness), and $s = 1/3$. That is, in the presence of a constant fraction of errors, the RLDC decoder outputs either the correct bit or a special error symbol $\bot$ with probability at least $2/3$. In the standard definition of an LDC, it is also common to set $s = 1/2 - \eps$ for a small constant $\eps$, meaning that in the presence of a constant fraction of errors, the LDC decoder outputs the correct bit with probability at least $1/2 + \eps$.

\paragraph{Our main results.}
Our first result shows that any linear $3$-RLDC with soundness error $1/2 - \eta$ is a $3$-LDC, and thus the threshold in \cref{ques:rldcequiv} must be at least $4$ for linear codes.
\begin{mtheorem}
\label{mthm:3rldcequiv}Let $\Code \colon \Bits^k \to \Bits^n$ be a linear  $(3, \delta, 1, \frac{1}{2} - \eta)$-RLDC with a possibly adaptive decoder. Then, for any $\eps > 0$, $\Code$ is a linear $(3, 2 \eta \delta \eps/3, 1, \eps)$-LDC.
In particular, if $\Code$ is a linear $(3, \Theta(1), 1, \frac{1}{3})$-RLDC, then $\Code$ is a linear $(3, \Theta(1), 1, \frac{1}{3})$-LDC. Furthermore, the same statement holds for RLCCs/LCCs.
\end{mtheorem}
By combining \cref{mthm:3rldcequiv} with the best known lower bounds for linear $3$-LDCs~\cite{AlrabiahGKM23,HsiehKMMS24,BasuHKL25,JanzerM25} and linear $3$-LCCs~\cite{AlrabiahG24}, we obtain the following new lower bounds for $3$-RLDCs and $3$-RLCCs.
\begin{corollary}
\label{cor:3rldclb}
Let $\Code \colon \Bits^k \to \Bits^n$ be a linear code. Then, the following hold:
\begin{enumerate}[(1)]
    \item 
If $\Code$ is a $(3, \delta, 1, \frac{1}{2} - \eps)$-RLDC with a possibly adaptive decoder, then $n \geq \Omega_{\delta, \eps}((k/\log k)^3)$.

\item
If $\Code$ is a $(3, \delta, 1, \frac{1}{2} - \eps)$-RLCC with a possibly adaptive decoder, then $n \geq 2^{\Omega_{\delta, \eps}(\sqrt{k/\log k})}$.
\end{enumerate}
\end{corollary}
For comparison, the prior best lower bound for $3$-RLDCs or $3$-RLCCs came from the works~\cite{GurL20,DallAgnolGL21,Goldreich23}, which prove a lower bound of $n \geq k^{1 + \Omega(1/q^2)}$ for any $q$-RLDC/RLCC, where the constant in the $\Omega(\cdot)$ is smaller than $1/10$. Thus, \cref{cor:3rldclb} gives a stronger lower bound for $3$-RLDCs, and a substantially stronger lower bound for $3$-RLCCs. 

The lower bounds in \cref{cor:3rldclb} require the RLDC/RLCC to have a soundness error of $1/2 - \eps$. However, for RLDCs/RLCCs (unlike LDCs/LCCs), any soundness error of $1 - \eps$ can be amplified to $1/2 - \eps$ via sequential repetition (see \cref{obs:expthreshold}), although this does increase the number of queries in the RLDC by a constant factor. In this way, one could view the requirement on the soundness error in \cref{cor:3rldclb} as a mild limitation, although we note that the lower bound for $2$-RLDCs in~\cite{BlockBCGLZZ23} also has the same requirement.

\cref{mthm:3rldcequiv} is a corollary of the following more general statement that we show. There is a soundness threshold $s(q)$, a function of $q$, such that \emph{any} linear $q$-RLDC with soundness error $s \leq (1 - \alpha)s(q)$ must be a linear $q$-LDC. That is, if a linear $q$-RLDC has soundness error better than $s(q)$, then the ``reason'' is that the $q$-RLDC is in fact a $q$-LDC. For $3$-RLDCs, this threshold $s(q)$ is $1/2$, which implies \cref{mthm:3rldcequiv}.
\begin{mtheorem}[Soundness error threshold for $q$-RLDCs; binary case of \cref{thm:formalrldcsoundnessthreshold}]
\label{mthm:rldcsoundnessthreshold}
    Let $s(q) \defeq 2^{-\floor{q/2}}$. Let $\Code \colon \Bits^k \to \Bits^n$ be a linear $(q,\delta, 1, s)$-RLDC with a possibly adaptive decoder, where $s \leq (1 - \alpha) s(q)$. Then, for any $\eps > 0$, $\Code$ is a linear $(q, \alpha \delta \eps/q, 1, \eps)$-LDC. Furthermore, the same result holds for RLCCs/LCCs.
\end{mtheorem}
\cref{mthm:rldcsoundnessthreshold} shows that the answer to \cref{ques:rldcequiv} depends on the soundness error $s$ that we choose in \cref{def:binary-rldc}.\footnote{Technically, to deduce that this depends on the soundness error one also needs to exhibit a $q$-RLDC (with soundness error $> s(q)$) that is also not a $q$-LDC. Existing constructions of RLDCs (\cite{BenSassonGHSV04,GurRR18,ChiesaGS20,AsadiS21,Goldreich24}) should satisfy this condition, and in this paper we give an RLDC with this property (\cref{mthm:rldcconstructions}).} The standard definition of RLDCs typically chooses $s = 1/3$ (see, e.g.,~\cite{BenSassonGHSV04,GurL20}), which is a constant smaller than $\frac{1}{2}$. If one wishes to have soundness error strictly less than, say, $s = 1/8$, then \cref{mthm:rldcsoundnessthreshold} implies that any $q$-RLDC that is additionally not a $q$-LDC must have $q \ge 8$.

More generally, \cref{mthm:rldcsoundnessthreshold} shows a qualitative difference between LDCs and RLDCs (that are not LDCs): the only way for a linear RLDC to have \emph{strong soundness}, a property satisfied by constant query linear LDCs, is for it to already be an LDC. Strong soundness is a natural property, typically desired in local testing or decoding algorithms, that intuitively says that corrupted codewords $y$ with fewer errors are decoded more successfully. More formally, strong soundness replaces Item (3) in \cref{def:binary-ldc,def:binary-rldc} with the following corresponding condition:
\begin{definition}[Strong soundness for LDCs and RLDCs]
\label{def:strongldc}
\phantom{w}
\begin{enumerate}[(1)]
\item ($\delta'$-strong LDC soundness) for all $b \in \Bits^k$, $i \in [k]$, and all $y \in \Bits^n$, $\Pr[\Dec^{y}(i) \neq b_i] \leq \Delta(y, \Code(b))/\delta'$.
\item ($\delta'$-strong RLDC soundness) for all $b \in \Bits^k$, $i \in [k]$, and all $y \in \Bits^n$, $\Pr[\Dec^{y}(i) \notin \{b_i, \bot\}] \leq \Delta(y, \Code(b))/\delta'$.
\end{enumerate}
\end{definition}
\cref{def:strongldc} says that the probability that the decoder is incorrect is proportional to the number of errors in $y$. The connection established in~\cite{KatzT00} between linear constant $q$-LDCs and ``smooth decoders'' implies that any $(q, \delta, 1, 1/2 - \eps)$-LDC (\cref{def:binary-ldc}) satisfies \cref{def:strongldc} for $\delta' = \delta/q$  (see \cref{app:strongsoundness}). On the other hand, \cref{mthm:rldcsoundnessthreshold} implies that no such analogous statement can hold for RLDCs unless all RLDCs are LDCs, which is known to be false (see, e.g., \cref{mthm:rldcconstructions}). Formally, we have the following corollary of~\cref{mthm:rldcsoundnessthreshold}.
\begin{corollary}[RLDCs with Strong Soundness are LDCs]
 Let $\Code \colon \Bits^k \to \Bits^n$ be a linear $q$-RLDC with perfect completeness and $\delta'$-strong soundness (\cref{def:strongldc}).
Then for any $\eps > 0$, $\Code$ is a linear $(q, s(q) \eps \delta'/4q, 1, \eps)$-LDC.
  Furthermore, the same result holds for RLCCs/LCCs.
 \end{corollary}

At first glance, the exponential dependence on $q$ in $s(q)$ in \cref{mthm:rldcsoundnessthreshold} may appear to be weak. However, a simple sequential repetition of the decoder, combined with a constant query RLDC that is not an $q$-LDC for any constant $q$, shows that the exponential dependence is necessary.
\begin{restatable}{observation}{expthreshold}
\label{obs:expthreshold}
Let $\Code \colon \Bits^k \to \Bits^n$ be a $(q, \delta, 1, \eps)$-RLDC with a possibly adaptive decoder. Then, for any integer $t \geq 1$, $\Code$ is a $(q t, \delta, 1, \eps^t)$-RLDC. The same statement holds for RLCCs.
\end{restatable}
\cref{obs:expthreshold} implies that if the threshold $s(q)$ in \cref{mthm:rldcsoundnessthreshold} did not decay exponentially with $q$, we could take any $(q, \delta, 1, \eps)$-RLDC (where $q, \delta, \eps$ are constant) and then choose $t$ to be a constant so that $\eps^t < s(qt)$ (which exists if $s(qt)$ does not decay exponentially), and this would show that the code is also a constant query LDC. However, as we will show (\cref{mthm:rldcconstructions}), there is a code that is an $O(1)$-RLDC but not a $q$-LDC for any constant $q$. We thus conclude that $s(q)$ must decay exponentially in $q$.

The proof of \cref{obs:expthreshold} is simple, and is included in \cref{sec:expthreshold}.
\medskip

\paragraph{Extensions of \cref{mthm:rldcsoundnessthreshold}.}
\cref{mthm:rldcsoundnessthreshold} makes two assumptions on the code: it assumes that $\Code$ is linear, and that it has perfect completeness. 
As we will discuss in \cref{sec:techsfooling}, our proof makes heavy use of the linearity of $\Code$, and it is not clear how to remove this assumption from \cref{mthm:rldcsoundnessthreshold}. However, we can extend \cref{mthm:rldcsoundnessthreshold} to codes with imperfect completeness via our next theorem, which shows how to convert any linear RLDC with imperfect completeness to a linear RLDC with perfect completeness.
\begin{mtheorem}[Binary case of \cref{thm:stronger-goldberg}]
\label{mthm:goldbergext}
Let $\Code \colon \Bits^k \to \Bits^n$ be a linear $(q, \delta, 1 - \eps, s)$-RLDC with a possibly adaptive decoder. Then, $\Code$ is a linear $(q, \delta, 1, s + 3 \eps)$-RLDC with a nonadaptive decoder. Furthermore, the same result holds for RLCCs/LCCs.
\end{mtheorem}
Combining \cref{mthm:rldcsoundnessthreshold,mthm:goldbergext}, we obtain the following corollary, which extends \cref{mthm:rldcsoundnessthreshold} to linear RLDCs with imperfect completeness.
\begin{corollary}
\label{cor:goldberg}
Let $s(q) \defeq 2^{-\floor{q/2}}$. Let $\Code \colon \Bits^k \to \Bits^n$ be a linear $(q,\delta, 1 - \eta, s)$-RLDC with a possibly adaptive decoder, where $s \leq (1 - \alpha) s(q) - 3\eta$. Then, for any $\eps > 0$, $\Code$ is a linear $(q, \alpha \delta \eps/q, 1, \eps)$-LDC. Furthermore, the same result holds for RLCCs/LCCs.
\end{corollary}
\cref{mthm:goldbergext} is a query-preserving version of a recent theorem of~\cite{Goldberg24}, which shows a similar result but requires an extra query. Namely,~\cite{Goldberg24} shows how to convert any linear $q$-RLDC with an adaptive decoder and imperfect completeness into a $(q+1)$-RLDC with a nonadaptive decoder and perfect completeness. This extra query made by~\cite{Goldberg24} is significant for us, as our soundness threshold $s(q) = 2^{-\floor{q/2}}$ satisfies $s(3) = 1/2$ but $s(4) = 1/4$, and so the result of~\cite{Goldberg24} only allows us to show that a $3$-RLDC (RLCC) with imperfect completeness and soundness error $1/4$ (which is smaller than the standard setting of $1/3$) is a $4$-LDC (LCC). For the case of RLCCs, this extra query is very significant, as the best lower bound for linear $4$-LCCs is only $n \geq \tilde{\Omega}(k^2)$~\cite{KerenidisW04}, whereas the best lower bound for (binary) linear $3$-LCCs is $n \geq 2^{\Omega(\sqrt{k/\log k})}$~\cite{AlrabiahG24}.

\parhead{Larger fields.} It is straightforward to extend~\cref{mthm:rldcsoundnessthreshold} to larger fields $\F$ by replacing the function $s(q)$ with $s_{\F}(q) \defeq \abs{\F}^{-\floor{q/2}}$, and we do this when we prove \cref{mthm:rldcsoundnessthreshold} in~\cref{sec:rldcsoundnessthreshold}. In \cref{app:2rldc-lower-bound}, we extend \cref{mthm:rldcsoundnessthreshold} to larger fields \emph{without} any field-dependent loss in $s(q)$ for the case of $q = 2$.

We prove \cref{mthm:3rldcequiv,mthm:rldcsoundnessthreshold} in \cref{sec:rldcsoundnessthreshold} and \cref{mthm:goldbergext} in \cref{sec:goldberg-strengthened}.

\paragraph{Constructions of RLDCs/RLCCs with small queries that are not LDCs.} To complement \cref{mthm:rldcsoundnessthreshold} and provide a partial answer to \cref{ques:rldcequiv,ques:rlccequiv}, we give a simple construction of RLDCs and RLCCs with small queries that are not constant query LDCs for any constant. To our knowledge, these are the first RLDC constructions to achieve constant queries for an explicit small constant.
\begin{mtheorem}[Constructions of constant query RLDCs/RLCCs that are not LDCs]
\label{mthm:rldcconstructions}
For every $k$, there is a linear code $\Code \colon \Bits^k \to \Bits^n$ where $n = k^{O(\log \log k)}$ such that $\Code$ is a $(15, \delta, 1, 1/3)$-RLDC and a $(41, \delta, 1, 1/2 - \eps)$-RLCC for a small constant $\eps > 0$. Furthermore, $\Code$ is not a $(O(\log k), O(n^{-1/3}), 1, 1/2 - \eps)$-LDC for any $\eps > 0$ (including subconstant $\eps$).
\end{mtheorem}
The blocklength of \cref{mthm:rldcconstructions} is worse than that of prior constructions, as it is slightly superpolynomial in $k$ rather than $\poly(k)$. However, the importance of \cref{mthm:rldcconstructions} is that (1) the number of queries made by the code is an explicit small constant, and (2) we can prove that the code is not a constant query LDC.
 Thus, when combined with \cref{cor:goldberg}, \cref{mthm:rldcconstructions} shows that the threshold in \cref{ques:rldcequiv} is somewhere between $4$ and $15$ for linear RLDCs with perfect completeness and soundness $1/2 - \eps$ for any constant $\eps > 0$: \cref{cor:goldberg} implies that linear $3$-RLDCs with soundness $1/2 - \eps$ are $3$-LDCs, and \cref{mthm:rldcconstructions} gives a $15$-RLDC with soundness $1/2 - \eps$ that is not an $q$-query LDC for any constant $q$.

While \cref{mthm:rldcconstructions} gives a construction of a $q$-RLDC with an explicit constant $q = 15$, one may wonder what the implicit constant $q$ is in previous works. We attempted to determine the number of queries $q$ needed for the RLDCs of~\cite{BenSassonGHSV04,GurRR18} to achieve soundness error $< 1/2$, and to the best of our knowledge, this $q$ is at least $10^7$ for both constructions. Intuitively, the reason the query complexity is so high is that these constructions make use of ``proof composition'' style results with robust probabilistically checkable proofs of proximity (PCPPs), and the soundness gap (defined as $1 - {}$soundness error) deteriorates multiplicatively with each composition step. This causes the soundness gap to deteriorate quickly even when just a few composition steps are used, and as a result the final soundness gap is smaller than $10^{-7}$. This means that the soundness error is at least $1 - 10^{-7}$, and so one must repeat the decoder many times (see \cref{obs:expthreshold}) to amplify the soundness error back down to $1/2$, which makes the query complexity large. In hindsight, the fact that these RLDCs use a large number of queries is perhaps unsurprising given that the constants in their proofs are likely not intentionally optimized.

We prove \cref{mthm:rldcconstructions} in \cref{sec:constructions}.


\section{Techniques}
\label{sec:techniques}
In this section, we give a proof overview of our main result, \cref{mthm:rldcsoundnessthreshold}. We will primarily focus on the case of $q = 2$ and $q = 3$ for RLDCs, though we shall also explain how to generalize the proof to arbitrary $q$ and also to RLCCs. For simplicity, we will assume that the decoder is nonadaptive; \cref{mthm:goldbergext} handles the case of adaptive decoders.
\subsection{The proof strategy for \cref{mthm:rldcsoundnessthreshold}}
\label{sec:proofstrategy}
Let $\Code \colon \Bits^k \to \Bits^n$ be a linear $(q, \delta, 1, s)$-RLDC with a nonadaptive decoder $\Dec(\cdot)$ satisfying the properties in \cref{def:binary-rldc} with $s \leq (1 - \alpha)s(q)$. Our goal is to show that $\Code$ is a $(q, \delta', 1, s')$-LDC for some $\delta', s'$. To do this, we need to construct a (possibly different) LDC decoder $\Dec'(\cdot)$ using the RLDC decoder $\Dec(\cdot)$. We will do this by ``opening up'' $\Dec(\cdot)$, i.e., we will crucially \emph{not} use $\Dec(\cdot)$ in a black-box manner.

\parhead{LDCs and smooth decoders.} The starting point of our proof is the following observation of~\cite{KatzT00}.
\begin{observation}
\label{obs:smoothtoldc}
Suppose that $\Code$ admits a $q$-query \emph{smooth decoder} $\Dec_{\mathrm{Smooth}}(\cdot)$, which is a nonadaptive local decoder with the following two properties:
\begin{enumerate}[(1)]
\item (perfect completeness) for every $b \in \Bits^k$ and $i \in [k]$, $\Pr[\Dec_{\mathrm{Smooth}}^{\Code(b)}(i) = b_i] = 1$, and
\item ($\eta$-smoothness) for every $i \in [k]$ and $j \in [n]$, $\Pr[\text{$\Dec_{\mathrm{Smooth}}(i)$ queries $j$}] \leq \frac{1}{\eta n}$.
\end{enumerate}
Then, for any $\eps > 0$, $\Code$ is a $(q, \eta \eps, 1, \eps)$-LDC.
\end{observation}
Indeed, \cref{obs:smoothtoldc} follows by a simple union bound, as if there are at most $\eta \eps n$ errors, then the probability that at least one of the $q$ queries made by the decoder is corrupted is at most $\frac{1}{\eta n} \cdot \eta \eps n = \eps$, and if all queries are uncorrupted then the decoder outputs $b_i$, by perfect completeness. Thus, to prove \cref{mthm:rldcsoundnessthreshold}, it suffices to extract a smooth decoder from $\Dec(\cdot)$.

\parhead{A canonical RLDC decoder.}
Let us now consider the behavior of $\Dec^{y}(i)$ for a fixed $i \in [k]$ and some $y \in \Bits^n$. Because $\Dec^y(i)$ is nonadaptive, we can view its operation as a two step process. First, $\Dec^y(i)$ samples a set of queries $Q$ of size $  \leq q$ (for simplicity, let us assume that $\abs{Q} = q$) from a query distribution $\cQ_i$ over ${[n] \choose \leq q}$ that does not depend on $y$. Then, it reads the values $y_j$ for each $j \in Q$, and outputs the value in $\{0,1, \bot\}$ of a (possibly randomized) function $f_Q(\{y_j\}_{j \in Q})$ that depends on the set of queries $Q$ and their answers $\{y_j\}_{j \in Q}$. We can now make the following simple observation: since the original decoder $\Dec(\cdot)$ has perfect completeness, there is a canonical choice of the decoding function $f_Q$ for each set $Q \subseteq [n]$. If the values of $y$ on $Q$ read by the decoder are consistent with some codeword $x = \Code(b)$, i.e., $y \vert_Q = x \vert_Q$, then the decoder must output $b_i$, the $i$-th bit of the underlying message in $x$, by perfect completeness. And, if the values of $y$ on $Q$ are inconsistent with \emph{all} codewords $x$, then $y$ must have an error in the set $Q$ and so the decoder can safely output $\bot$, as this can only decrease the soundness error of the decoder.

\parhead{Decomposing the canonical RLDC decoder into smooth and nonsmooth parts.}
The above observation thus implies that we can view the decoder $\Dec^{y}(i)$ as being solely determined by the query distribution $\cQ_i$. Given a distribution $\cQ_i$, we can define the set of ``nonsmooth'' or ``heavy'' queries as $H_i \defeq \{j \in [n] : \Pr_{Q \gets \cQ_i}[j \in Q] > \frac{q}{\delta n}\}$. Note that for a $(\delta/q)$-smooth decoder, the set $H_i$ is empty. We clearly have $\abs{H_i} \leq \delta n$, as
\begin{flalign*}
q = \sum_{j \in [n]} \Pr_{Q \gets \cQ_i}[j \in Q] \geq \sum_{j \in H_i}\Pr_{Q \gets \cQ_i}[j \in Q] \geq \frac{q\abs{H_i}}{\delta n} \mper
\end{flalign*}
In particular, this means that for any $b \in \Bits^k$, the RLDC decoder $\Dec(i)$ on input $i$ must satisfy $\Pr[\Dec^{y}(i) \notin \{b_i, \bot\}] \leq s$ for any $y$ where $y_j = \Code(b)_j$ for all $j \notin H_i$. That is, if we only introduce errors in the ``heavy set'' $H_i$, we have introduced a small enough fraction of errors so that the soundness condition of the RLDC decoder still applies.

Call a set $Q \in {[n] \choose q}$ ``$i$-smoothable'' if one can recover $b_i$ from $x \vert_{Q \setminus H_i}$ for any codeword $x = \Code(b)$, and let $p_{i,\textrm{good}} = \Pr_{Q \gets \cQ_i}[\text{$Q$ is $i$-smoothable}]$. Let $\cQ_{i,\textrm{good}}$ be the distribution $\cQ_i$ conditioned on the output $Q$ being ``$i$-smoothable'',
and let $\cQ_{i,\textrm{bad}}$ be $\cQ_i$ conditioned on $Q$ being not ``$i$-smoothable''. Recall that because there is an optimal canonical decoder, a (nonadaptive) decoder is determined completely by its query distribution. Thus, the set of query
distributions $\cQ_{i,\textrm{good}}$ for each $i \in [k]$ defines a decoder that we call $\Dec_{\mathrm{LDC}}$, and similarly $\{\cQ_{i,\textrm{bad}}\}_{i \in [k]}$ defines a decoder $\Dec_{\mathrm{RLDC}}$. Furthermore, for any $y \in \Bits^n$ and any $i \in [k]$, for every output $\sigma \in \{0,1,\bot\}$, it holds that
\begin{equation*}
\Pr[\Dec^y(i) = \sigma] = p_{i,\textrm{good}}\Pr[\Dec_{\mathrm{LDC}}^y(i) = \sigma] +  (1 - p_{i,\textrm{good}})\Pr[\Dec_{\mathrm{RLDC}}^y(i) = \sigma] \mper
\end{equation*}
In other words, we can view the behavior of $\Dec^y(i)$ as follows: with probability $p_{i,\textrm{good}}$, it runs $\Dec_{\mathrm{LDC}}^y(i)$, and with probability $1 - p_{i,\textrm{good}}$ it runs $\Dec_{\mathrm{RLDC}}^y(i)$.
The two decoders $\Dec_{\mathrm{LDC}}$ and $\Dec_{\mathrm{RLDC}}$ have their names chosen to indicate that  $\Dec_{\mathrm{LDC}}$ is the ``LDC part'' or ``smooth part'' of the original decoder $\Dec$, and $\Dec_{\mathrm{RLDC}}$ is the ``pure RLDC part'' or ``nonsmooth part'' of the original decoder.

\parhead{Showing that the ``smooth part'' is nontrivial by breaking soundness of the ``nonsmooth'' part.} To finish proving that $\Code$ is an LDC, it remains to argue that $p_{i,\textrm{good}} \geq \Omega(1)$ for each $i \in [k]$, i.e., the ``smooth part'' of the RLDC decoder is nontrivially large. This is because the decoder $\Dec_{\mathrm{LDC}}$ is $\delta p_{i,\textrm{good}}/q$-smooth\footnote{Here, we sample $Q$ from $\cQ_{i, \textrm{good}}$ and only query $Q \setminus H_i$. Note that because $Q$ is $i$-smoothable, we can still decode $b_i$ from $Q \setminus H_i$, and so we have preserved perfect completeness while making the decoder smooth.} for each $i \in [k]$, and so by \cref{obs:smoothtoldc} it is an LDC decoder that can tolerate $\delta p/q$ errors, where $p = \min_{i \in [k]} p_{i, \textrm{good}}$. We thus want $p_{i,\textrm{good}} \geq \Omega(1)$ for each $i \in [k]$, so that $\delta p/q = \Omega(\delta/q) = \Omega(1)$ is at least a constant.

To argue that $p_{i,\textrm{good}} \geq \Omega(1)$ for each $i \in [k]$, we will fix $i \in [k]$ and we show that any decoder that makes at least one ``nonsmooth'' query has soundness error at least $s(q) = 2^{-\floor{q/2}}$. More formally, we show that for each $b \in \Bits^k$, there is a $y \in \Bits^n$ where $y_v = \Code(b)_v$ for $v \notin H_i$ such that $\Pr[\Dec_{\mathrm{RLDC}}^y(i) \notin \{b_i, \bot\}] \geq s(q)$. That is, $y$ agrees with a codeword on the ``smooth queries'' and only disagrees on some of the ``nonsmooth queries'', which also implies that $\Delta(y, \Code(b)) \leq \abs{H_i} \leq \delta n$. Note that by soundness of the original decoder $\Dec$, this implies
\begin{flalign*}
&(1 - \alpha) s(q)\geq s \geq \Pr[\Dec^y(i) \notin \{b_i, \bot\}] \\
&= p_{i,\textrm{good}}\Pr[\Dec_{\mathrm{LDC}}^y(i) \notin \{b_i, \bot\}] +  (1 - p_{i,\textrm{good}})\Pr[\Dec_{\mathrm{RLDC}}^y(i) \notin \{b_i, \bot\}] \\
&\geq s(q)(1 - p_{i,\textrm{good}}) \mcom
\end{flalign*}
and so $p_{i,\textrm{good}} \geq \alpha$.

We explain how to break soundness of $\Dec_{\mathrm{RLDC}}$ in the next section.
\subsection{Analyzing the ``nonsmooth'' linear RLDC decoder}
\label{sec:techsfooling}
It remains to show that $\Dec_{\mathrm{RLDC}}$ has soundness error at least $s(q) = 2^{-\floor{q/2}}$. As explained above, we will show that for each $b \in \Bits^k$, there exists $y \in \Bits^n$ that agrees with $x = \Code(b)$ on all coordinates in $[n] \setminus H_i$ such that $\Pr[\Dec_{\mathrm{RLDC}}^{y}(i) = 1 - b_i] \geq s(q)$. Because the code is linear, without loss of generality we may assume that $b = 0^k$, so that $x = 0^n$, and our goal is to show the existence of such a $y \in \Bits^n$ where $\Pr[\Dec_{\mathrm{RLDC}}^{y}(i) = 1] \geq s(q)$.

\parhead{Fooling a fixed set $Q$.}
Let us now explain how to construct such a $y$. Fix a set $Q$ in the support of the query distribution of $\Dec_{\mathrm{RLDC}}$. Because we have an optimal canonical decoder, when $\Dec_{\mathrm{RLDC}}$ queries $Q$, it simply checks if  $y \vert_Q = x \vert_Q$ for some codeword $x$, and if so, then it must output the bit $b_i$ where $x = \Code(b)$.

As a first attempt, suppose we choose $y$ so that $y \vert_{H_i}$ is random and otherwise $y$ agrees with $0^n$, the original uncorrupted codeword. Then, with probability at least $2^{-q}$, it holds that $y \vert_Q = x \vert_Q$ for some codeword $x = \Code(b)$ with $b_i = 1$. Indeed, because $Q$ is not ``$i$-smoothable'', the message bit $b_i$ cannot be recovered from $x \vert_{Q \setminus H_i}$ for a codeword $x = \Code(b)$, and so there must exist a codeword $x = \Code(b)$ with $b_i = 1$ such that $x \vert_{Q \setminus H_i}$ agrees with the codeword $0^n$. And, since $y \vert_{H_i}$ is random, we have that $y \vert_{Q \cap H_i} = x \vert_{Q \cap H_i}$ with probability at least $2^{-\abs{Q \cap H_i}} \geq 2^{-q}$. We note that this observation is sufficient to prove \cref{mthm:rldcsoundnessthreshold} with $s(q) = 2^{-q}$.

To prove \cref{mthm:rldcsoundnessthreshold} with $s(q) = 2^{-\floor{q/2}}$, however, we require a more sophisticated analysis that will end up using the linearity of $\Code$ quite strongly. For a linear code $\Code$, $y \vert_Q$ agrees with a codeword $x$ on $Q$ if and only if $y \vert_Q$ satisfies a certain system of linear equations that are the ``local parity check constraints on $Q$''. For example, it could be the case that $Q = \{j_1,j_2,j_3\}$ where any codeword $x$ satisfies $x_{j_1} + x_{j_2} + x_{j_3} = 0$, and there are no other local constraints. In this case, if $y$ satisfies $y_{j_1} + y_{j_2} + y_{j_3} = 0$, then there exists some codeword $x$ where $x_{j_1} = y_{j_1}$, $x_{j_2} = y_{j_2}$, and $x_{j_3} = y_{j_3}$.

However, the above observation is not enough for us, as we additionally need to make $\Dec_{\mathrm{RLDC}}$ output $1$. To make this happen, we will again make use of the linear structure of $\Code$. Because $\Code \colon \Bits^k \to \Bits^n$ is a linear map, for every $j \in [n]$ there exists $v_j \in \Bits^k$ such that for every $b \in \Bits^k$, $\Code(b)_j \defeq \ip{b, v_j}$; the vector $v_j$ is simply the $j$-th row of the generator matrix of $\Code$. Note that ``local constraints'' above correspond to linear dependencies among the rows of the generator matrix. That is, $x_{j_1} + x_{j_2} + x_{j_3} = 0$ for all codewords $x$ if and only if $v_{j_1} + v_{j_2} + v_{j_3} = 0^k$.

The fact that $\Dec_{\mathrm{RLDC}}$ has perfect completeness implies that one can recover $b_i$ exactly from $x = \Code(b)$ restricted to $Q$. In linear algebraic terms, this means that the $i$-th standard basis vector $e_i$ is in $\vspan(\{v_j : j \in Q\})$. We have assumed that $Q$ is not ``$i$-smoothable'', meaning that one \emph{cannot} recover $b_i$ from $x \vert_{Q \setminus H_i}$, as otherwise $Q$ is a set that ``belongs to'' $\Dec_{\mathrm{LDC}}$. So, $e_i$ is not in the span of $\vspan(\{v_j : j \in Q \setminus H_i\})$.

Our main technical lemma (\cref{lem:fooling}) shows that if we pick $y$ such that: (1) $y_j = \Code(0^k)_j = 0$ for $j \notin H_i$, and (2) $y_j = \Code(b)_j$ for $j \in H_i$ where $b \in \Bits^k$ is \emph{random} with $b_i = 1$, then with probability at least $s(q) = 2^{-\floor{q/2}}$, $y \vert_Q = x \vert_Q$ for some codeword $x = \Code(b')$ where $b'_i = 1$. That is, $y \vert_Q$ satisfies all local checks in $Q$ and is consistent with a codeword that has $i$-th message bit equal to $1$, and thus when this occurs $\Dec_{\mathrm{RLDC}}$ outputs $1$.

\begin{remark}
\label{rem:rldcweakness}
The intuition for our main technical lemma (and indeed the main intuition behind \cref{mthm:rldcsoundnessthreshold}) is as follows. Firstly, in order for the RLDC decoder to be ``not LDC-like'', it needs to make queries to the nonsmooth set $H_i$. Secondly, in order for the RLDC decoder to be hard to fool, it must use the queries to $Q \setminus H_i$ to check  consistency with the queries $Q \cap H_i$; if there are no consistency checks, then we can freely choose the values of $y$ on $Q \cap H_i$ and easily fool the decoder. Finally, when $q$ is small (say $q = 2$ or $3$), the decoder cannot simultaneously make many queries to both $H_i$ and $[n] \setminus H_i$, i.e., one of $Q \cap H_i$ or $Q \setminus H_i$ must be small, and this is an inherent weakness that allows us to fool $\Dec_{\mathrm{RLDC}}$ with reasonable probability.
\end{remark}

\parhead{Casework for $q = 2$.}
Let us now explain a simple argument to prove our main technical lemma when $q = 2$. Let $Q = \{j_1,j_2\}$. We split the analysis into $3$ cases, depending on $\abs{Q \cap H_i}$, i.e., the number of ``nonsmooth queries'' made in $Q$. As we shall see, the case of $q = 2$ is simple and also somewhat degenerate, in that it is not possible to have a nontrivial ``local check''. In fact, this is the reason that we can show a very good lower bound for $2$-RLDCs over large fields (see \cref{app:2rldc-lower-bound}).
\begin{enumerate}[(1)]
\item Case 1: $\abs{Q \cap H_i} = 0$. In this case, $Q \cap H_i = \emptyset$, and therefore $e_i$ is in $\vspan(\{v_j\}_{j \in Q \setminus H_i})$, contradicting the fact that $Q$ is not $i$-smoothable. Thus, this case cannot occur.

\item Case 2: $\abs{Q \cap H_i} = 1$. In this case, $Q = \{j_1,j_2\}$ where $j_1 \in H_i$ and $j_2 \notin H_i$. We have that $e_i \notin \vspan(v_{j_2})$, which implies that $v_{j_2} \ne e_i$. We also have $e_i \in \vspan(v_{j_1},v_{j_2})$, which implies that either $e_i = v_{j_1}$, or $e_i = v_{j_1} + v_{j_2}$.

We also have at most one ``local check'': either $v_{j_1} + v_{j_2} = 0^k$, or else there are no local check constraints. However, if there is a local check constraint, then if $v_{j_1} = e_i$, this implies that $v_{j_2}= e_i$, and so $e_i \in \vspan(v_{j_2})$ (a contradiction), and if $v_{j_1} + v_{j_2} = e_i$, then the local check constraint implies $e_i = 0^k$ (a contradiction). Thus, we cannot have a local check constraint.

Because $j_2 \notin H_i$, we have $y_{j_2} = 0$, and so we fool the decoder if and only if $y_{j_1} = 1$. This clearly happens with probability at least $1/2$, as either $v_{j_1} = e_i$, in which case $y_{j_1} = 1$ with probability $1$, or else $v_{j_2} \ne e_i$ and is nonzero, in which case $y_{j_1} = 1$ with probability $1/2$.

\item Case 3: $\abs{Q \cap H_i} = 2$. In this case, $y \vert_Q = x \vert_Q$ for some $x = \Code(b')$ where $b'_i = 1$, by definition. Thus, the canonical decoder outputs $1$ with probability $1$.
\end{enumerate}
In all cases, we see that the decoder outputs $1$ with probability at least $s(2) = 1/2$ over the random choice of $y$, as required.

\parhead{Generalizing to higher $q$.}
The above casework proves the desired claim for $q = 2$. One can repeat a similar case-by-case analysis for $q = 3$, but the analysis quickly becomes unwieldy as $q$ increases. For example, when $q = 3$ one can have $Q = \{j_1,j_2,j_3\}$ where $j_1,j_2 \in H_i$, $j_3 \notin H_i$, and $v_{j_1} + v_{j_2} = e_i$ and $v_{j_2} + v_{j_3} = 0$. That is, we can have both a ``decoding constraint'' that sums to $e_i$ and a ``check constraint'' that sums to $0$. Again, one can verify that our choice of random $y$ will satisfy both these constraints with probability $1/2$: this is because the two constraints imply $v_{j_1} + v_{j_2} = e_i$, and this constraint is satisfied with probability $1$ (because $y$ is consistent with $x = \Code(b')$ with $b'_i = 1$ on indices in $H_i$), and the constraint $v_{j_2} + v_{j_3} = 0$ is satisfied with probability $1/2$, independently of the first constraint. We note that it is easy to show how to set the values $y_j$ for $j \in H_i$ to fool the decoder for a particular $Q$, but the key difficulty is we need to find a single global $y$ that fools an $s(q)$-fraction of the $Q$'s simultaneously.

To find a single global $y$, we show that if we sample $y$ from the distribution as above, i.e., where $y \vert_{H_i}$ is chosen to be $\Code(b) \vert_{H_i}$ where $b$ is uniformly random with $b_i = 1$, then for any set $Q$ that is not $i$-smoothable, with probability at least $s(q)$ over the draw of $y$, $y$ fools the decoder when it queries $Q$. Recall that $y$ fools the decoder if and only if $y \vert_Q \in \Code' \vert_Q$, where $\Code'$ is the affine subspace $\{\Code(b) \vert_{Q \cap H_i} : b_i = 1\}$. This requires checking that \begin{inparaenum}[(1)] \item $y \vert_{Q \cap H_i}$ is in $\Code' \vert_{Q \cap H_i}$, \item $y \vert_{Q \setminus H_i}$ is in $\Code' \vert_{Q \setminus H_i}$, and \item $y \vert_{Q \cap H_i}$ is consistent with $0^{Q \setminus H_i}$ and the constraint that $b_i = 1$\end{inparaenum}. Notice that $y$ satisfies all of the type (1) constraints with probability $1$, since $y \vert_{Q \cap H_i}$ is drawn uniformly over that set, and $y$ satisfies all of the type (2) constraints by definition because $Q$ is not $i$-smoothable. We can view the type (3) constraints as imposing a system of inhomogeneous linear constraints on the vector $y \vert_{Q \cap H_i}$, so that $y$ fools the decoder if and only if $y \vert_{Q \cap H_i}$ lies in a particular affine subspace.

Formally, we show that for each $Q$ and each codeword $\Code(b)$, there is a vector $z \in \Bits^{Q \cap H_i}$ and linear subspaces $\cV$ and $\cW$ in $\Bits^{Q \cap H_i}$ with $\cV \subseteq \cW$ and $\dim(\cW/\cV) \leq \min(\abs{Q \cap H_i}, \abs{Q \setminus H_i})$ such that $y \vert_Q$ fools the decoder if and only if $y \vert_{Q \cap H_i} - z$ lies in $\cW^{\perp}$. Intuitively, the vector $z$ shifts the affine subspaces to be linear subspaces, and then the subspace $\cW$ is the set of ``local parity check constraints'' on $Q \cap H_i$ satisfied by all codewords $\Code(b')$ with $b'_i = 0$ that are also $0$ on $Q \setminus H_i$ (type (1) and type (3) constraints), and the subspace $\cV$ is the set of constraints on $Q \cap H_i$ satisfied by all codewords $\Code(b')$ with $b'_i = 0$ (type (1) constraints only). The distribution of $y \vert_{Q \cap H_i} - z$ is then uniform over the larger subspace $\cV^{\perp}$ in $\Bits^{Q \cap H_i}$ that contains $\cW^{\perp}$. Hence, $y$ lies in the desired affine subspace with probability at least $2^{-\dim(\cW/\cV)}$ and this is at least $2^{-\floor{q/2}}$, as one can show that $\dim(\cW/\cV) \leq \min(\abs{Q \cap H_i}, \abs{Q \setminus H_i})\leq \floor{q/2}$. This is the ``inherent weakness'' mentioned in \cref{rem:rldcweakness}: one of $\abs{Q \cap H_i}$ or $\abs{Q \setminus H_i}$ must have size $\leq \floor{q/2}$, and they control the dimension of the ``extra local checks'' that $y \vert_{Q \cap H_i}$ must satisfy in order to fool the decoder.

The full proof of \cref{mthm:rldcsoundnessthreshold} is in \cref{sec:rldcsoundnessthreshold}.

\parhead{Generalizing to RLCCs.} The above analysis considers the behavior of $\Dec^y(i)$ on each input $i$ separately, and shows how to convert it to a smooth decoder. Thus, it seamlessly generalizes to RLCCs, as we can consider the behavior of the RLCC decoder $\Dec^y(u)$ on each input $u$ separately as well, and convert $\Dec^y(u)$ to a smooth decoder for each $u$ to obtain an LCC.


\section{Preliminaries}
\subsection{Basic notation}
We let $[n]$ denote the set $\{1, \dots, n\}$. For a natural number $t \in \N$, we let ${[n] \choose t}$ be the collection of subsets of $[n]$ of size exactly $t$.

Given a string $x \in \Sigma^n$ and a set $S \subseteq [n]$, we define $x\rvert_S$ to be the restriction of $x$ to the indices in $S$.
Similarly, for a set of strings $X \subseteq \Sigma^n$, we define $X\rvert_S \coloneqq \{x\rvert_S : x \in X\}$. If $\Sigma$ is an alphabet with a distinguished element $0 \in \Sigma$, we also define $X_{\subseteq S} \defeq \{x \in X: \supp(x) \subseteq S\}$, where $\supp(x) \defeq \{i \in S: x_i \ne 0\}$.

Given a finite field $\F$ and $x,y \in \F^n$, we let $\ip{x,y} = \sum_{i = 1}^n x_i y_i$ denote their inner product.

\begin{definition}[Hamming distance]
    For strings $x,y \in \Sigma^n$, the (absolute) \emph{Hamming distance} $\Delta(x,y)$ is the number of indices where $x$ differs from $y$.
    The \emph{relative Hamming distance} $\bar{\Delta}(x,y) \coloneqq \Delta(x,y)/n$ is the fraction of indices where $x$ differs from $y$.
    \begin{equation*}
    \Delta(x,y)
    \coloneqq
    \#\{i : x_i \neq y_i\},
    \qquad \bar{\Delta}(x,y) \defeq \Delta(x,y)/n\mper
    \end{equation*}
    If $Y \subseteq \Sigma^n$, then $\Delta(x,Y)$ denotes the minimum Hamming distance between $x$ and an element $y \in Y$:
    \begin{equation*}
    \Delta(x,Y) \defeq \min_{y \in Y} \Delta(x,y)\mper
    \end{equation*}
\end{definition}
\subsection{Linear codes}
\begin{definition}[Linear codes]
\label{def:linear-code}
    A \emph{linear code} $\Code$ over a finite field $\F$ is an injective $\F$-linear map $\Code \colon \F^k \to \F^n$. The integer $k$ is the length of the message of the code, and the integer $n$ is the blocklength of the code. Because $\Code$ is injective, $k = \dim(\im(\Code))$ is the dimension of the code.

    We will sometimes specify $\Code$ as a dimension $k$ linear subspace of $\F^n$, rather than by its encoding map, i.e., we have $\Code \subseteq \F^n$, and we write $x \in \Code$ to indicate that $x$ is in the image of the (implicitly defined) linear map from $\F^k \to \Code\subseteq \F^n$.
    
    A linear code $\Code$ can be described by both a ``generator matrix'' and a ``parity check matrix.''
    \begin{itemize}
        \item (Generator matrix) Given a linear code $\Code \colon \F^k \to \F^n$, there is a matrix $M \in \F^{n \times k}$ such that $\Code(\msg) = A\msg$ for each $\msg \in \F^k$. The matrix $A$ is called the ``generator matrix'' of the code.

        \item (Parity check matrix) Given a linear code $\Code \colon \F^k \to \F^n$, there is a matrix $B \in \F^{(n-k) \times n}$ such that $x \in \im(\Code)$ if and only if $B x = 0^{n - k}$. The matrix $B$ is called a ``parity check matrix'' for $\Code$.
    \end{itemize}
    We say that a family of linear codes is \emph{explicit} if there is a uniform efficient algorithm for computing generator matrices or parity check matrices for the family.
\end{definition}
\begin{definition}[Dual code]
\label{def:dualcode}
Let $\Code \colon \F^k \to \F^n$. Its \emph{dual code}, denoted by $\Code^{\perp}$ is the linear subspace of $\F^n$ given by $\Code^{\perp} = \{y \in \F^n : \ip{x,y} = 0 \ \forall x \in \Code\}$. Here, $\ip{x,y}= \sum_{i = 1}^n x_i y_i$ is the standard inner product.
\end{definition}

The following fact relates the dual code of $\Code \vert_S$ to the dual code of $\Code$.
\begin{fact}
\label{fact:restriction-support}
    Let $\Code \subseteq \F^n$ be a linear code, and let $S \subseteq [n]$.
    Then, $(\Code\rvert_S)^\perp \defeq  \Code^{\perp}_{\subseteq S}$, where $\Code^{\perp}_{\subseteq S} = \{y \in \Code^{\perp} : \supp(y) := \{i : y_i \ne 0\} \subseteq S\}$.
\end{fact}
\begin{proof}
We have that $z \in (\Code\rvert_S)^\perp\subseteq\F_2^S$ if and only if
\begin{equation*}
\forall x\in \Code, ~0 = \ip{x \vert_S, z} = \sum_{i \in S} x_i z_i = \sum_{i \in S} x_i z_i + \sum_{i \notin S} x_i \cdot 0 = \ip{x,y} \mcom
\end{equation*}
where $y_i = z_i$ for $i \in S$ and $y_i = 0$ for $i \notin S$, and the latter statement is true if and only if $y \in \Code^{\perp}_{\subseteq S}$, as desired.
\end{proof}

The following fact describes exactly when a linear combination $\ip{v, \msg}$ of the message $\msg$ is determined by $\Code(b) \vert_Q$ for a set $Q$.
\begin{fact}
  \label{fact:indep-rows}
Let $\Code \colon \F^k \to \F^n$ be a linear code. For each $j \in [n]$, let $v_j$ denote the $j$-th row of the generator matrix, so that $\ip{v_j, \msg} = \Code(\msg)_j$ for all $b \in \F^k$. 

Let $Q \subseteq [n]$ and let $x^* = \Code(\msg^*)$. Let $v \in \F^k$. Then, every $x = \Code(\msg)$ with $x \vert_Q = x^* \vert_Q$ satisfies $\ip{v, \msg} = \ip{v, \msg^*}$ if and only if $v\in \vspan(\{v_j\}_{j \in Q})$. Furthermore, if $v\notin \vspan(\{v_j\}_{j \in Q})$, then for every $\sigma \in \F$, there exists $\msg \in \F^k$ such that $x \vert_Q = x^* \vert_Q$ and $\ip{v, \msg} = \sigma$.

That is, if we are given the values of a codeword $x^*$ restricted to $Q$, it ``fixes'' some linear combination of the message symbols if and only if the linear combination is in the span of rows of the generator matrix corresponding to the set $Q$, and otherwise it is ``free''.
\end{fact}

\subsection{Locally decodable/correctable codes and their relaxed notions}
Below, we define LDCs/RLDCs and LCCs/RLCCs.
\begin{definition}[Locally decodable codes]
  \label{def:ldc}
A code $\Code \colon \Sigma^k \to \Sigma^n$ is a $(q, \delta, c, s)$-LDC if there exists a randomized decoding algorithm $\Dec(\cdot)$ with the following properties. The algorithm $\Dec(\cdot)$ is given oracle access to a string $y \in \Sigma^n$, takes an index $i \in [k]$ as input, and outputs a symbol in $\Sigma$ with the following guarantees:
\begin{enumerate}[(1)]
\item ($q$-queries) for any $y$ and $i$, the algorithm $\Dec^{y}(i)$ reads at most $q$ indices of $y$,
\item ($c$-completeness) for all $b \in \Sigma^k$ and $i \in [k]$, $\Pr[\Dec^{\Code(b)}(i) = b_i] \geq c$, and
\item ($(\delta, s)$-soundness error) for all $b \in \Sigma^k$, $i \in [k]$, and all $y \in \Sigma^n$ with $\Delta(y, \Code(b)) \leq \delta n$, $\Pr[\Dec^{y}(i) \ne b_i] \leq s$.
\end{enumerate}
\end{definition}

\begin{definition}[Locally correctable codes]
\label{def:lcc}
A code $\Code \colon \Sigma^k \to \Sigma^n$ is a $(q, \delta, c, s)$-LCC if there exists a randomized decoding algorithm $\Dec(\cdot)$ with the following properties. The algorithm $\Dec(\cdot)$ is given oracle access to a string $y \in \Sigma^n$, takes an index $u \in [n]$ as input, and outputs a symbol in $\Sigma$ with the following guarantees:
\begin{enumerate}[(1)]
\item ($q$-queries) for any $y$ and $u$, the algorithm $\Dec^{y}(u)$ reads at most $q$ indices of $y$,
\item ($c$-completeness) for all $b \in \Sigma^k$ and $u \in [n]$, $\Pr[\Dec^{\Code(b)}(u) = \Code(b)_u] \geq c$, and
\item ($(\delta, s)$-soundness error) for all $b \in \Sigma^k$, $u \in [n]$, and all $y \in \Sigma^n$ with $\Delta(y, \Code(b)) \leq \delta n$, $\Pr[\Dec^{y}(u) \ne \Code(b)_u] \leq s$.
\end{enumerate}
\end{definition} 

We will now define smooth decoders/correctors. These are decoders which only need to work on valid codewords, but must not favor querying any one index heavily.

\begin{definition}[Smooth decoder]
\label{def:smoothldc}
An $\eta$-\emph{smooth decoder} of a code $\Code\colon \Sigma^k\to\Sigma^n$ is a decoder $\Dec(\cdot)$ such that \begin{enumerate}[(1)]
    \item ($q$-queries) for any $y$ and $i\in [k]$, the algorithm $\Dec^y(i)$ reads at most $q$ indices of $y$,
    \item (perfect completeness) for every $\msg\in \Sigma^k$ and $i\in [k]$, $\Pr[\Dec^{\Code(b)}(i) = b_i] = 1$, and 
    \item ($\eta$-smoothness) for every $i\in [k]$ and $j\in [n]$, $\Pr[\Dec(i)\text{ queries }j]\le \frac{1}{\eta n}$.
\end{enumerate}
    
\end{definition}

\begin{definition}[Smooth corrector]
\label{def:smoothlcc}
A $\eta$-\emph{smooth corrector} of a code $\Code\subset \Sigma^n$ is a decoder $\Dec(\cdot)$ such that \begin{enumerate}[(1)]
    \item ($q$-queries) for any $y$ and $u\in [n]$, the algorithm $\Dec^y(u)$ reads at most $q$ indices of $y$,
    \item (perfect completeness) for every $\msg\in \Sigma^k$ and $u\in [n]$, $\Pr[\Dec^{\Code(b)}(u) = \Code(b)_u] = 1$, and 
    \item ($\delta$-smoothness) for every $u,j \in [n]$, $\Pr[\text{$\Dec(u)$ queries $j$}]\le \frac{1}{\eta n}$.
\end{enumerate}
\end{definition}

The notion of smooth decoders was introduced in~\cite{KatzT00} because of their equivalence to locally decodable codes. For this paper, we will need the following simple fact.
\begin{fact}[Smooth decoder implies local decoding]
\label{fact:smoothtoldc}
    Let $\Code$ be a code with a $q$-query $\eta$-smooth decoder (corrector). Then, $\Code$ is a $(q,\eta\eps,1,\eps)$-LDC (LCC).
\end{fact}

\begin{proof}
    We will prove the smooth decoder is indeed the desired local decoder (the corrector case follows analogously). Perfect completeness follows by definition in \Cref{def:smoothldc}, so it remains to show that the soundness error is at most $\eps$. Consider a received word $y$ such that $\Delta(y, C(b))\le \eta\eps n$ for some $b\in \Sigma^k$. Let $S\subset[n]$ be the indices where $y$ differs from $C(b)$. By $\eta$-smoothness, we have that for any $v\in S$, $\Pr[\text{$\Dec(i)$ queries $v$}]\le \frac{1}{\eta n}$. Union bounding over all $v\in S$, it follows
    \begin{equation*}
    \Pr[\text{$\Dec(i)$ does not query $S$}] \ge 1-\frac{1}{\eta n} \cdot |S| \geq 1 - \eps \mper
    \end{equation*}
    Now if $\Dec^y(i)$ never queries any index in $S$, its local view is consistent with the codeword $C(b)$. Consequently, the decoder must output $b_i$ with probability $1$ by perfect completeness. Hence, $\Dec(\cdot)$ has soundness error $\leq \eps$ as desired.  
\end{proof}

\begin{definition}[Relaxed locally decodable codes]
  \label{def:rldc}
A code $\Code \colon \Sigma^k \to \Sigma^n$ is  a $(q, \delta, c, s)$-RLDC if there exists a randomized decoding algorithm $\Dec(\cdot)$ with the following properties. The algorithm $\Dec(\cdot)$ is given oracle access to a string $y \in \Sigma^n$, takes an index $i \in [k]$ as input, and outputs either a symbol in $\Sigma$ or a special symbol $\bot$ with the following guarantees:
\begin{enumerate}[(1)]
\item ($q$-queries) for any $y$ and $i$, the algorithm $\Dec^{y}(i)$ reads at most $q$ indices of $y$,
\item ($c$-completeness) for all $b \in \Sigma^k$ and $i \in [k]$, $\Pr[\Dec^{\Code(b)}(i) = b_i] \geq c$, and
\item ($(\delta, s)$-relaxed soundness error) for all $b \in \Sigma^k$, $i \in [k]$, and all $y \in \Sigma^n$ with $\Delta(y, \Code(b)) \leq \delta n$, $\Pr[\Dec^{y}(i) \notin \{b_i, \bot\}] \leq s$.
\end{enumerate}
\end{definition}
\begin{definition}[Relaxed locally correctable codes]
\label{def:rlcc}
A code $\Code \colon \Sigma^k \to \Sigma^n$ is  a $(q, \delta, c, s)$-RLDC if there exists a randomized decoding algorithm $\Dec(\cdot)$ with the following properties. The algorithm $\Dec(\cdot)$ is given oracle access to a string $y \in \Sigma^n$, takes an index $u \in [n]$ as input, and outputs either a symbol in $\Sigma$ or a special symbol $\bot$ with the following guarantees:
\begin{enumerate}[(1)]
\item ($q$-queries) for any $y$ and $u$, the algorithm $\Dec^{y}(u)$ reads at most $q$ indices of $y$
\item ($c$-completeness) for all $b \in \Sigma^k$ and $u \in [n]$, $\Pr[\Dec^{\Code(b)}(u) = \Code(b)_u] \geq c$,
\item ($(\delta, s)$-relaxed soundness error) for all $b \in \Sigma^k$, $u \in [n]$, and all $y \in \Sigma^n$ with $\Delta(y, \Code(b)) \leq \delta n$, $\Pr[\Dec^{y}(u) \notin \{\Code(b)_u, \bot\}] \leq s$.
\end{enumerate}
\end{definition}

For RLDCs/RLCCs with perfect completeness, we can assume that the decoder behaves in a certain ``canonical'' way.
\begin{fact}[Canonical behavior of a local decoder]
\label{fact:canonical}
Let $\Code \colon \Bits^k \to \Bits^n$ be a $(q, \delta, 1, s)$-RLDC with decoder $\Dec_1$. Then, there is a decoder $\Dec_2$ for $\Code$ with perfect completeness such that (1) $\Code$ is a $(q, \delta, 1, s)$-RLDC using $\Dec_2$, and (2) whenever $\Dec_2^{y}(i)$ queries a set $Q$, its behavior is as follows:
\begin{enumerate}[(1)]
\item Find $x = \Code(b)$ such that $x \vert_Q = y \vert_Q$. If there is no such $x$, output $\bot$.
\item Otherwise, output $b_i$.
\end{enumerate}
Furthermore, if $\Dec_1$ is nonadaptive, then so is $\Dec_2$, and the analogous statement also holds for RLCCs.

We say that the RLDC decoder of $\Code$ is ``canonical'' if it follows the operation of $\Dec_2$.
\end{fact}
\begin{proof}
We define $\Dec_2$ by (1) running $\Dec_1$ until it has finished making all of its queries, and then (2) following the above decoding behavior. Note that $\Dec_2$ has perfect completeness by definition, as $\Dec_1$ has perfect completeness. The only difference between $\Dec_1$ and $\Dec_2$ is that $\Dec_2$ may output $\bot$ when $\Dec_1$ outputs some other symbol. But, this can only decrease the soundness error, which finishes the proof.
\end{proof}

In the case of linear codes, the canonical decoder described in \cref{fact:canonical} has a nice linear algebraic structure, coming from the dual code.

\begin{fact}
\label{fact:canon-linear}
    Let $\Code \colon \F^k \to \F^n$ be a linear $(q, \delta, 1, s)$-RLDC with a canonical decoder $\Dec$. Let $Q \subseteq [n]$ be a subset, and let $z \in \F^Q$. The behavior of the $\Dec$, when it queries the set $Q$ and sees the local view $z$, is described as follows:
    \begin{enumerate}[(1)]
\item Check if $z \in \Code \vert_Q$, and output $\bot$ otherwise;
\item Let $j_1, \dots, j_t \in Q$ be such that $x_{j_1} + \dots + x_{j_t} = b_i$ for all $x = \Code(b)$ (such a set must exist by perfect completeness). Output $z_{j_1} + \dots + z_{j_t}$.
    \end{enumerate}
    Furthermore, condition (1) can be checked by verifying that $M z = 0$ for some matrix $M \in \F^{t \times Q}$, where $t \leq \abs{Q}$. We call these constraints the ``testing constraints'' and the constraint in Item (2) the ``decoding constraint''.
\end{fact}

\begin{proof}
Item (1) exactly matches the behavior of the decoder in~\cref{fact:canonical}. To see Item (2), we use the generator matrix definition of the map $\Code$. That is, for each $j \in [n]$, there exists $v_j \in \F^k$ such that $x_j = \ip{v_j, b}$ when $x = \Code(b)$. With this perspective, we can recover $b_i$ from $\{x_{j}\}_{j \in Q}$ if and only if $e_i \in \vspan(\{v_j\}_{j \in Q})$. Hence, by perfect completeness, $e_i$ must be in the span, and so there exist $j_1, \dots, j_t \in Q$ such that $v_{j_1} + \dots + v_{j_t} = e_i$.

To prove the ``furthermore'', we observe that since $\Code \vert_Q$ is a linear subspace, we can check membership via a system of homogeneous linear equations, which yields the matrix $M$.
\end{proof}

\subsection{Proof of \cref{obs:expthreshold}}
\label{sec:expthreshold}
We prove \cref{obs:expthreshold}, which is restated below.
\expthreshold*
\begin{proof}
Consider the new RLDC decoder that runs the original decoder $t$ times independently, and outputs a bit $\sigma \in \Bits$ if all invocations of the decoder output $\sigma$, and otherwise the decoder outputs $\bot$. This decoder clearly satisfies perfect completeness, and has soundness error at most $\eps^t$ because the $t$ invocations of the decoder are independent.
\end{proof}
\subsection{Linearity testing}
We recall the well-known result for linearity testing over $\F_2$.
\begin{fact}[Linearity Test~\cite{BlumLR93,BellareCHKS95}]
\label{fact:linearitytest}
Let $G \colon \F_2^n \to \F_2$ be an arbitrary function, and let $\bar{\Delta}(G, \LIN)$ be the minimum, over linear functions $F \colon \F_2^n \to \F_2$, of $\E_{x \in \F_2^n}[G(x) \ne F(x)]$. Then, $\Pr_{x,y \in \F_2^n}[G(x) + G(y) + G(x+y) = 0] \leq 1 - \bar{\Delta}(G)$. That is, if the linearity test passes with probability at least $1 - \eps$, then $\bar{\Delta}(G, \LIN) \leq \eps$.
\end{fact}

We also recall that one can self-correct functions that are close to linear.
\begin{fact}[Self-correction of near-linear functions]
\label{fact:linrecovery}
Let $G \colon \F_2^n \to \F_2$ be an arbitrary function, and let $F \colon \F_2^n \to \F_2$ be a linear function. Let $\bar{\Delta}(F,G) = \E_{x \in \F_2^n}[G(x) \ne F(x)]$. Then, for any $x \in \F_2^n$, $\Pr_{y \in \F_2^n}[G(x+y) + G(y) = F(x)] \geq 1 - 2\bar{\Delta}(F,G)$.
\end{fact}

\section{Relaxed Locally Decodable Codes Cannot Have Strong Soundness}
\label{sec:rldcsoundnessthreshold}
In this section, we prove \cref{mthm:rldcsoundnessthreshold}, which shows a qualitative difference between linear RLDCs and LDCs: an LDC has strong soundness (soundness error can be made arbitrarily low by adjusting the decoding radius; see \cref{app:strongsoundness}), while an RLDC does not unless it is also an LDC. We will in fact prove the following theorem, which is a generalization of \cref{mthm:rldcsoundnessthreshold} to any finite field for the case of nonadaptive decoders. The case of adaptive decoders and imperfect completeness is handled generically by \cref{thm:stronger-goldberg}, which we prove in \cref{sec:goldberg-strengthened}.
\begin{theorem}[\cref{mthm:rldcsoundnessthreshold} for nonadaptive decoders and any field]
\label{thm:formalrldcsoundnessthreshold}
    Let $s_{\F}(q) \defeq \abs{\F}^{-\floor{q/2}}$. Let $\Code \colon \F^k \to \F^n$ be a linear $(q,\delta, 1, s)$-RLDC with a nonadaptive decoder, where $s \leq (1 - \alpha) s_{\F}(q)$. Then, for any $\eps > 0$, $\Code$ is a linear $(q, \alpha \delta \eps/q, 1, \eps)$-LDC. Furthermore, the same result holds for RLCCs/LCCs.
\end{theorem}

\begin{proof}
Let $\Code \colon \F^k \to \F^n$ be a linear $(q,\delta, 1, s)$-RLDC with $s \leq (1 - \alpha) s_{\F}(q)$. We will follow the proof outline from \cref{sec:techniques}.

Let $\Dec(\cdot)$ be the nonadaptive RLDC decoder of $\Code$. Because $\Dec(\cdot)$ is nonadaptive, we may assume (\cref{fact:canonical}) that it behaves as follows. For each $i$, there is a distribution $\cQ_i$ over subsets of $[n]$ of size at most $q$, and the decoder $\Dec^y(i)$ on input $i$ simply samples $Q \gets \cQ_i$, reads $y_v$ for each $v \in Q$, and then follows the behavior of the canonical decoder in~\cref{fact:canonical}.

\parhead{Defining the smooth decoder $\Dec_{\mathrm{LDC}}(\cdot)$.} Let $i \in [k]$ be a message index. For each $i \in [k]$, we partition the codeword indices $[n]$ into heavy (``nonsmooth'') and light (``smooth'') indices as follows.
\begin{gather*}
  H_i \coloneq \{j\in[n]: \Pr_{Q\gets \cQ_i} [j\in Q] > q/\delta n\}\\
  L_i \coloneq \{j\in[n]: \Pr_{Q\gets \cQ_i} [j\in Q] \leq q/\delta n\}
\end{gather*}
We then have
\begin{equation*}
q \geq \E_{Q\gets \cQ^{(i)}}\Big[\sum_{j\in [n]} \1(j\in Q)\Big] \ge \sum_{j\in H_i} \E_{Q\gets \cQ^{(i)}}\left[\1(j\in Q)\right]\ge \frac{q|H_i|}{\delta n} \mper
\end{equation*}
Hence, $|H_i|\le \delta n$.

We now split $\cQ_i$ into the ``smooth part'' and the ``nonsmooth part''. Recall from the definition of a linear code (\cref{def:linear-code}) that for each $j \in [n]$, there exists $v_j \in \F^k$ such that for every $b \in \F^k$, $\Code(b)_j = \ip{v_j, b}$. We now define the notion of a ``smoothable'' query set, which are sets $Q$ where one can recover $b_i$ from only the light indices, i.e., $Q \cap L_i$.
\begin{definition}
\label{def:smoothableset}
For a set $Q$ in the support of $\cQ_i$, we call $Q$ \emph{$i$-smoothable} if $e_i \in \vspan\{v_j : j \in Q \cap L_i\}$.
\end{definition}
Let us now define the following query distribution $\tilde{\cQ}_i$. In the distribution $\tilde{\cQ}_i$, we sample $Q \gets \cQ_i$ conditioned on $Q$ being $i$-smoothable (we will show that this probability is nonzero, so this is well-defined), and then we output $\tilde{Q} \defeq Q \cap L_i$. Note that $\tilde{Q}$ is $i$-smoothable since $Q$ is.

Given the distribution $\tilde{\cQ}_i$ for each $i \in [k]$, we define a decoder $\Dec_{\mathrm{LDC}}(\cdot)$ as follows. On input $i$, $\Dec_{\mathrm{LDC}}^y(i)$ draws $\tilde{Q} \gets \tilde{\cQ}$, and then uses the ``decoding constraint'' to decode $b_i$. That is, because $\tilde{Q}$ is $i$-smoothable, there exists a subset $T \subseteq \tilde{Q}$ such that $\sum_{j \in T} v_j = e_i$, and the decoder outputs $\sum_{j \in T} y_j$.

\parhead{Arguing smoothness of $\Dec_{\mathrm{LDC}}(\cdot)$.} We will now show that $\Dec_{\mathrm{LDC}}(\cdot)$ is $(\alpha \delta/q)$-smooth (\cref{def:smoothldc}).
For each $i \in [k]$, let $p_{i, \mathrm{good}}$ be the probability that $Q \gets \cQ_i$ is $i$-smoothable. Let $\Dec_{\mathrm{RLDC}}(\cdot)$ denote the decoder that (1) samples $Q \gets \cQ_i$ conditioned on $Q$ being \emph{not} $i$-smoothable, and then (2) decodes using the canonical decoder. Observe that we can view the original decoder as simply calling $\Dec_{\mathrm{RLDC}}(\cdot)$ with probability $1 - p_{i, \mathrm{good}}$, where $i \in [k]$ is the input index (and, with probability $p_{i, \mathrm{good}}$, the decoder does something else). We then have that for any $b \in \F^k$ and $y \in \F^n$ with $\Delta(y, \Code(b)) \leq \delta n$ and any $i \in [k]$, it holds that
\begin{equation}
\label{eq:deducesmooth}
(1 - \alpha) s_{\F}(q) \geq s \geq \Pr[\Dec^y(i) = \sigma \in \F \setminus \{b_i\}] \geq (1 - p_{i, \mathrm{good}})\Pr[\Dec_{\mathrm{RLDC}}^y(i) = \sigma \in \F \setminus \{b_i\}] \mper
\end{equation}
Our main technical lemma is the following lemma, which we will use to lower bound the soundness error of $\Dec_{\mathrm{RLDC}}$.
\begin{lemma}[Fooling a set $Q$]
\label{lem:fooling}
Let $\Code \colon \F^k \to \F^n$ be a linear code. Let $Q = H \cup L$ where $H$ and $L$ are disjoint. Let $v^* \in \F^k$, and suppose that $v^* \notin \vspan(\{v_j : j \in L\})$. Let $b \in \F^k$, and let $x = \Code(b)$. Fix $i \in [k]$, and let $\sigma \in \F$.

Let $b' \gets \F^k$ be sampled uniformly at random with $\ip{v^*, b'} = \sigma$, and let $x' = \Code(b')$. Define $z \in \F^Q$ to be $z_j = x_j$ for $j \in L$ and $z_j = x'_j$ for $j \in H$. Then, with probability at least $\abs{\F}^{-\min(\abs{H}, \abs{L})}$ over the choice of $b'$, there exists $b'' \in \F^k$ such that $\Code(b'') \vert_Q = z$ and $\ip{v^*, b''} = \sigma$.
\end{lemma}
\cref{lem:fooling} says the following. Suppose we are given a query set $Q$ and a codeword $x = \Code(b)$, and we want to corrupt $x \vert_Q$ by only modifying its values on $H_i$ so that the corrupted version of $x \vert_Q$ is now equal to $x'' \vert_Q$ for a different codeword $x''$. Then, if we corrupt $x \vert_Q$ by replacing $x \vert_{Q \cap H_i}$ with a uniformly random codeword, then it will satisfy this condition with some nontrivial probability. Furthermore, if we additionally wish the codeword $x''$ to be equal to $\Code(b'')$ for some $b''$ satisfying a particular inhomogeneous linear constraint $\ip{v^*, b''} = \sigma$, then this is possible provided that $v^* \notin \vspan(\{v_j : j \in Q \cap L_i\})$ and $v^* \in \vspan(\{v_j : j \in Q\})$.

We postpone the proof of \cref{lem:fooling} to \cref{sec:fooling}, and for now use it to finish the proof of \cref{thm:formalrldcsoundnessthreshold}. As we shall shortly see, \cref{lem:fooling} implies that the soundness error of $\Dec_{\mathrm{RLDC}}$ is at least $s_{\F}(q)$. 

Indeed, let $v^* = e_i$, and let $Q$ be any set in the support of $\cQ_i$ that is not $i$-smoothable. Let $b \in \F^k$ and let $x = \Code(b)$. Suppose that we define (a distribution over) $y \in \F^n$ with $\Delta(y, \Code(b)) \leq \delta n$ by (1) sampling $b' \gets \F^k$ with $b'_i \ne b_i$, and (2) setting $y$ to be $y \vert_{H_i} = \Code(b') \vert_{H_i}$, and $y \vert_{L_i} = \Code(b) \vert_{L_i}$. We clearly have that $\Delta(y, \Code(b)) \leq \abs{H_i} \leq \delta n$ holds with probability $1$. On the other hand, by \cref{lem:fooling}, with probability at least $\abs{\F}^{-\min(\abs{Q \cap H_i}, \abs{Q \cap L_i})} \geq \abs{\F}^{-\floor{q/2}} = s_{\F}(q)$, it holds that $y \vert_Q = x'' \vert_Q$ for some $x'' = \Code(b'')$ where $b''_i = b'_i \ne b_i$. Because $\Dec_{\mathrm{RLDC}}$ behaves as the canonical decoder, it must therefore output $b''_i \ne b_i$.

The above shows that for any $i \in [k]$, there is a distribution over $y$ with $\Delta(y, \Code(b)) \leq \delta n$ such that $\E_{y}[\Pr[\Dec_{\mathrm{RLDC}}^{y}(i) \ne b_i]] \geq s_{\F}(q)$. Hence, by averaging, there exists $y$ such that this holds. This implies that the soundness error of $\Dec_{\mathrm{RLDC}}^{y}(i)$ is at least $s_{\F}(q)$, and so by \cref{eq:deducesmooth}, we conclude that $(1 - \alpha) s_{\F}(q) \geq (1 - p_{i, \mathrm{good}})s_{\F}(q)$, which implies that $p_{i, \mathrm{good}} \geq \alpha$, and this holds for all $i \in [k]$.

With this lower bound on $p_{i, \mathrm{good}}$ in hand, let us now argue smoothness of $\Dec_{\mathrm{LDC}}(\cdot)$. Indeed, for any $j \in [n]$, we have that 
\begin{equation*}
\Pr[\textrm{$\Dec_{\mathrm{LDC}}^y(i)$ queries $j$}] = \Pr_{\tilde{Q} \gets \tilde{\cQ_i}}[j \in \tilde{Q}] = \Pr_{Q \gets \cQ_i}[j \in Q \cap L_i \given \textrm{$Q$ is $i$-smoothable}] \mper
\end{equation*}
Notice that this probability is $0$ if $j \in H_i$, and otherwise for $j \in L_i$ we have
\begin{align*}
\Pr[\textrm{$\Dec_{\mathrm{LDC}}^y(i)$ queries $j$}] & = \Pr_{Q \gets \cQ_i}[j \in Q \given \textrm{$Q$ is $i$-smoothable}]  \\ & = \frac{\Pr_{Q \gets \cQ_i}[j \in Q \wedge \textrm{$Q$ is $i$-smoothable}]}{\Pr_{Q \gets \cQ_i}[\textrm{$Q$ is $i$-smoothable}]} \\
&= \frac{\Pr_{Q \gets \cQ_i}[j \in Q \wedge \textrm{$Q$ is $i$-smoothable}]}{p_{i, \mathrm{good}}} \\ &  \leq \frac{\Pr_{Q \gets \cQ_i}[j \in Q]}{{p_{i, \mathrm{good}}}} \\ &  \leq \frac{q}{\alpha\delta n} \mcom
\end{align*}
where we use that $j \in L_i$. Hence, $\Dec_{\mathrm{LDC}}$ is $(\alpha \delta/q)$-smooth. By \cref{fact:smoothtoldc}, this implies that $\Code$ is a $(q, \alpha \delta \eps/q, 1, \eps)$-LDC with decoder $\Dec_{\mathrm{LDC}}$, which proves \cref{thm:formalrldcsoundnessthreshold} for the case of RLDCs.

\parhead{Extension to the RLCC case.} Let us now briefly explain how to extend the above proof to the case of RLCCs. As we are proceeding by analyzing the behavior of $\Dec(\cdot)$ on each input, this generalizes seamlessly to RLCCs. The only difference in the proof is that in \cref{lem:fooling}, we take $v^* = v_j$, where $v_j$ is the $j$-th row of the generator matrix of $\Code$ and $j \in [n]$ is the input to $\Dec(\cdot)$. That is, $v^*$ is no longer necessarily a standard basis vector $e_i$. All the remaining steps of the proof are unchanged.
\end{proof}

\subsection{Proof of \cref{lem:fooling}}
\label{sec:fooling}
In this subsection, we prove \cref{lem:fooling}.
\begin{proof}[Proof of \cref{lem:fooling}]
Let $Q \subseteq [n]$, and let $Q = H \cup L$ where $H \cap L = \emptyset$. Let $b \in \F^k$, and let $x = \Code(b)$. Fix $i \in [k]$, and let $\sigma \in \F$. Our goal is to show that, if we choose $b' \gets \F^k$ uniformly at random with $\ip{v^*, b'} = \sigma$, then with probability at least $\abs{\F}^{-\min(\abs{H}, \abs{L})}$, the string $z \in \F^Q$ defined as $z_j = x_j$ for $j \in L$ and $z_j = x'_j$ for $j \in H$ is consistent with some other codeword $\Code(b'')$ that also satisfies $\ip{v^*, b''} = \sigma$.

Because $\Code$ is linear, without loss of generality we may assume that $b = 0^k$, so that $z_j = 0$ for all $j \in L$. We wish to argue that $\Pr_{z}[\exists b'' \text{ s.t. } z = \Code(b'') \vert_Q \wedge \ip{v^*, b''} = \sigma] \geq \abs{\F}^{-\min(\abs{H}, \abs{L})}$, when $z$ is drawn from the distribution defined in \cref{lem:fooling}.

Let us first argue that such a $z$ exists. This implies that the probability is at least $\abs{\F}^{-\abs{Q}}$, which is already sufficent to prove \cref{thm:formalrldcsoundnessthreshold} with $s_{\F}(q) = \abs{\F}^{-q}$.

Indeed, because $v^* \notin \vspan\{v_j : j \in L\}$ and there is a codeword that is identically $0$ on all $j \in L$, it follows by \cref{fact:indep-rows} that such a $z$ exists. Let $z^* = \Code(b^*) \vert_Q$ be any such solution, with corresponding codeword $b^*$.

Because there exists such a $b^*$, by the linearity of $\Code$, it suffices to show that $\Pr_{z \gets \cD}[\exists b'' \text{ s.t. } z = \Code(b'') \vert_Q \wedge \ip{v^*, b''} = 0] \geq \abs{\F}^{-\min(\abs{H}, \abs{L})}$, where $\cD$ is the distribution over $z$ given by (1) sampling $b' \gets \F^k$ uniformly at random with $\ip{v^*, b'} = 0$, and (2) outputting $z \in \F^Q$ where $z_j = 0$ for $j \in L$ and $z_j = \Code(b')_j$ for $j \in H$. This is because any $z$ satisfies the above condition if and only if $z + z^*$ satisfies the conditions in \cref{lem:fooling}.

Let us now define the following vector spaces: $\cV_0 = \{\Code(b') : \ip{v^*, b'} = 0\}$, $\cW_0 = \{\Code(b') : \ip{v^*, b'} = 0, \Code(b')_j = 0 \ \forall j\in L\}$. Note that $\cW_0 \subseteq \cV_0$, and these are both linear subspaces. We let $\cV_0^{\perp}$ and $\cW_0^{\perp}$ denote their corresponding dual subspaces (\cref{def:dualcode}).

By \cref{fact:restriction-support}, $(\cV_0^{\perp})_{\subseteq H}$ are the set of local constraints defining $\cV_0 \vert_H$, and similarly for $(\cW_0^{\perp})_{\subseteq H}$ and $\cW_0 \vert_H$. We will refer to these subspaces often, and so we let $\cV \defeq (\cV_0^{\perp})_{\subseteq H}$ and $\cW \defeq (\cW_0^{\perp})_{\subseteq H}$.

A key fact that we will use is that $((\cV_0^{\perp})_{\subseteq Q}) \vert_H = \cW$. This is immediate by definition, as $(\cV_0^{\perp})_{\subseteq Q}$ is the set of local constraints that defines $\cV_0 \vert_Q$, and when we enforce $C(b')_j = 0$ for all $j \in L$, the restriction of any constraint in $\cV_0 \vert_Q$ to the set $H$ is a local constraint in $\cW_0$ on the subset $H = Q \setminus L$. For notational convenience, we will let $\cU = (\cV_0^{\perp})_{\subseteq Q}$, so that $\cU \vert_H = \cW$, and $\cU_{\subseteq H} = \cV$.

We now show the following two claims, which together imply \cref{lem:fooling}.
\begin{claim}
\label{claim:problowerbound}
$\Pr_{z \gets \cD}[\exists b'' \text{ s.t. } z = \Code(b'')\vert_Q \wedge \ip{v^*, b''} = 0] \geq \abs{\F}^{-\dim(\cW/\cV)}$.
\end{claim}

\begin{claim}
\label{claim:dimupperbound}
$\dim(\cW/\cV) \leq \min(\abs{H}, \abs{L})$.
\end{claim}
Indeed, \cref{claim:problowerbound,claim:dimupperbound} imply that $\Pr_{z \gets \cD}[\exists b'' \text{ s.t. } z = \Code(b'')\vert_Q \wedge \ip{v^*, b''} = 0] \geq \abs{\F}^{-\min(\abs{H}, \abs{L})}$, which we have already shown suffices to prove \cref{lem:fooling}.
\end{proof}

\begin{proof}[Proof of \cref{claim:problowerbound}]
By definition of the subspace $\cV_0$, we have that $w \defeq z \vert_H$ is uniformly distributed over $\cV_0 \vert_H$. By definition of the subspace $\cV_0$, we have that $z$ satisfies the desired condition if and only if $w \in \cW_0$. Thus, $\Pr_{w \gets \cV_0 \vert_H}[w \in \cW_0 \vert_H] \geq \abs{\F}^{-t}$ where $t$ is the number of ``independent checks'' in $\cW$ that are not in $\cV$. We have that $t = \dim(\cW/\cV)$, which gives us the claim.

More formally, let $r^{(1)}, \dots, r^{(d)}$ be vectors in $\cW$ that are linearly independent in $\cW/\cV$ (and hence also linearly independent in $\cW$). By definition of $\cW/\cV$, any $r \in \cW$ can be expressed as $s + \sum_{j \in T} r^{(j)}$, where $s \in \cV$. Now, $w$ satisfies $\ip{w,s} = 0$, and hence if $\ip{w,r^{(j)}} = 0$ holds for all $j \in [d]$, then $\ip{w,r} = 0$ for all $r \in \cW$. Finally, observe that because $r^{(1)}, \dots, r^{(t)}$ are linearly independent in $\cW/\cV$, the elements $\ip{w,r^{(j)}}$ are independent and uniformly random from $\F$ when $w \gets V_0$. Hence, the probability that $\ip{w,r^{(j)}} = 0$ for all $j \in [t]$ is at least $\abs{\F}^{-t}$.
\end{proof}

\begin{proof}[Proof of \cref{claim:dimupperbound}]
Because $\cV, \cW$ are subspaces in $\F^H$, it follows that $\dim(\cW/\cV) \leq \dim(\cW) \leq \abs{H}$. Thus, it remains to prove that $\dim(\cW/\cV) \leq \abs{L}$, which is the nontrivial case.

This proof uses the following key fact that we established earlier: $\cU \vert_H = \cW$ and $\cU_{\subseteq H} = \cV$, where $\cU \subseteq \F^Q$ is a subspace.

Suppose that $\dim(\cW/\cV) \geq \abs{L} + 1$. Let $d = \abs{L} + 1$, and let $r^{(1)}, \dots, r^{(d)}$ be elements of $\cW$ that are linearly independent in $\cW/\cV$. Because $\cW = \cU \vert_H$, for each $j \in [t]$, there exist $s^{(1)}, \dots, s^{(d)} \in \F^Q$ with $\supp(s^{(j)}) \subseteq L$ such that $r^{(j)} + s^{(j)} \in \cU$ and $\supp(r^{(j)} + s^{(j)}) \subseteq H$. Now, because $s^{(1)}, \dots, s^{(d)}$ are in $\F^Q$ and have support contained in $L$, they lie in a subspace of dimension at most $\abs{L}$, and since $d = \abs{L} + 1$, they must be linearly dependent. Hence, there exist $\alpha_1, \dots \alpha_d \in \F$, not all zero, such that $\sum_{j = 1}^d \alpha_j s^{(j)} = 0$ in $\F^Q$. We then have that 
\begin{equation*}
\sum_{j = 1}^d \alpha_j r^{(j)} = \sum_{j = 1}^d \alpha_j (r^{(j)} + s^{(j)}) -  \sum_{j = 1}^d \alpha_j  s^{(j)} = \sum_{j = 1}^d \alpha_j (r^{(j)} + s^{(j)}) \mper
\end{equation*}
Now, because $\supp(r^{(j)} + s^{(j)}) \subseteq H$ for all $j$, this implies that $\supp(\sum_{j = 1}^d \alpha_j (r^{(j)} + s^{(j)})) \subseteq H$ also, and therefore $\sum_{j = 1}^t \alpha_j (r^{(j)} + s^{(j)})$ is an element of $\cU$ whose support is contained in $H$, i.e., an element of $\cU_{\subseteq H} = \cV$. Hence, $\sum_{j = 1}^d \alpha_j (r^{(j)} + s^{(j)}) \in \cV$. It thus follows that $\sum_{j = 1}^d \alpha_j (r^{(j)} + s^{(j)})$ is $0$ in $\cW/\cV$, and hence $\sum_{j = 1}^d \alpha_j r^{(j)}$ is also $0$ in $\cW /\cV$. Therefore, $r^{(1)}, \dots, r^{(d)}$ are linearly independent in $\cW/\cV$, which proves the claim.
\end{proof}


\section{Query-Preserving Goldberg Transformation}
\label{sec:goldberg-strengthened}
Up to this point, we have assumed that our RLDC or RLCC has a local
decoder/corrector which has \emph{perfect completeness} (always
returns the right answer for a valid codeword)
and which is \emph{nonadaptive} (the local view is sampled before any
queries have been made).
We may assume that such a decoder has a \emph{canonical} behavior,
as we showed in \cref{fact:canonical},
and we rely on this structure in our proofs.
However, what if we begin with an RLDC or RLCC with a
local decoder that has imperfect completeness, adaptivity, or both?
Goldberg's transformation \cite{Goldberg24} shows that such a decoder can be
transformed (potentially inefficiently) into a nonadaptive decoder
with perfect completeness:
\begin{theorem}[{\cite{Goldberg24}}]
  Every linear RLDC or RLCC\footnote{\cite{Goldberg24} only proves this theorem for binary codes, but the proof easily extends to all
  finite fields.} has a nonadaptive decoder with perfect completeness:
  \begin{enumerate}
  \item If $\Code\colon \F^k \to \F^n$ is a linear systematic $(q, \delta, 1-\epsilon, s)$-RLDC,
    then $\Code$ is also a $(q+1, \delta, 1, s+\epsilon)$-RLDC with a nonadaptive
    decoder.
  \item If $\Code\colon \F^k \to \F^n$ is a linear $(q, \delta, 1-\epsilon, s)$-RLCC,
    then $\Code$ is also a $(q+1, \delta, 1, s+\epsilon)$-RLCC with a nonadaptive decoder.
  \end{enumerate}
\end{theorem}
We can use this theorem to lift our lower bound to general linear RLDCs and
RLCCs.
However, the extra query (from $q$ to $q+1$) is very costly for our
purposes.
In particular,
it would imply that the lower bound in
\cref{cor:3rldclb} only applies to \emph{2-RLDCs}
with imperfect completeness,
rather than 3-RLDCs.
As discussed in \cref{sec:intro},
this extra query has a substantial impact on our results. To avoid losing this query, we give an modified analysis of the main result of~\cite{Goldberg24} that does not lose this additional query. In doing so, we will lose slightly in the soundness error. We also do not need $\Code$ to be
systematic, which was required in \cite{Goldberg24}.
\begin{theorem}[General form of \cref{mthm:goldbergext}]
  \label{thm:stronger-goldberg}
  If $\Code\colon \F^k \to \F^n$ is a linear $(q, \delta, 1-\epsilon, s)$-RLDC,
  then $\Code$ is also a $(q, \delta, 1, s+(2 + 1/(\abs{\F} - 1)) \eps)$-RLDC with a nonadaptive
  decoder. The same result holds for RLCCs.
\end{theorem}

We will rely on the lemmas and proofs from \cite{Goldberg24}
and point out where we differ.
To begin, we can model any adaptive randomized relaxed local
decoder\footnote{This entire section will be written in terms of
  RLDCs, but the same proof works for RLCCs.} $\Dec(i)$ as a distribution over decision trees $\Gamma \leftarrow \mathcal{D}(i)$,
where vertices are labeled with query indices from $[n]$
and edges are labeled with the symbols read from those query indices.
Then, each leaf $\ell$ in each tree $\Gamma$ corresponds to
an ordered query tuple $Q$ and a corresponding string $\sigma \in \F^Q$.
Every leaf $\ell$ is labeled with a symbol from $\F \cup \{\bot\}$,
which is the value returned by the decoder.
With this in mind, the Goldberg transformation is based around
relabeling leaves and rerandomizing the input.

\subsection{Relabeling leaves}
We first modify the decoder for each $i \in [k]$
to shift the input by
a uniformly chosen codeword $\Code(\tilde{\msg})$
(and then subtract $\tilde{\msg}_i$ from the answer if it is not $\bot$);
this gives a new decoder $\Dec_{R}$ with the same parameters
as the initial decoder.
Next,
every decision tree leaf
which contributes to the completeness error
needs to be relabeled.
For each $i \in [k]$,
these are the leaves in trees $\Gamma \in \supp(\mathcal{D}(i))$
which are reached by some uncorrupted codeword $\Code(\msg)$
but which are \emph{not} labeled with $\msg_i$.
Goldberg shows that changing the label of this leaf to $\msg_i$
trades completeness error for soundness error,
and we can iterate this process to get
an adaptive relaxed local decoder with perfect completeness.
The input rerandomization is crucial to make this lemma work.
\begin{lemma}[leaf relabeling, {\cite[Claim 18]{Goldberg24}}]
  \label{lem:leaf-relabeling}
  Let $\Dec_R(i)$ be the rerandomized decoder for index $i$,
  and let $\Dec'_R(i)$ be the same decoder
  with a single leaf of a single decision tree relabeled as described.
  Then, if $\Dec_R(i)$ has completeness error $\epsilon$
  and soundness error $s$,
  and $\Dec'_R(i)$ has completeness error $\epsilon'$
  and soundness error $s'$, then
  \begin{equation*}
    s' - s \le \epsilon - \epsilon'\mper
  \end{equation*}
\end{lemma}

There is a critical aspect to this relabeling step that
we have so far overlooked.
What if, for some $i \in [k]$,
there is a leaf $\ell$
corresponding to queried values $(Q, \sigma)$
where the $i$-th message index is
\emph{linearly independent}
of the indices in $Q$ (\cref{fact:indep-rows})?
This means that the decoder receives absolutely
no information about the $i$-th message index
from its queries, \emph{even}
if all of the queries match an uncorrupted codeword.
Thus, there is no fixed label we can
give to this leaf to reduce completeness error.
Call these leaves \emph{toxic}.
\begin{definition}[Toxic leaves]
  A leaf $(Q, \sigma)$ of a decision tree $\Gamma$
  in the support of $\mathcal{D}(i)$
  is \emph{toxic} if
  there exist codewords $\Code(\msg), \Code(\msg')$
  such that $\Code(\msg)\rvert_Q = \Code(\msg')\rvert_Q = \sigma$,
  but $\msg_i \neq \msg'_i$.
  By \cref{fact:indep-rows},
  the local view $Q$
  does not give any information on the $i$-th index --- for any
  local view and any codeword,
  there are $\abs{\F}-1$ other codewords,
  one for each possible symbol,
  which look identical on the local view yet differ on
  the $i$-th message index.
  On the other hand,
  if a leaf is non-toxic, then by \cref{fact:indep-rows} the $i$-th message index for any
  codeword is completely determined by its restriction to $Q$.
\end{definition}

\subsection{Isolating toxic leaves}
Goldberg disambiguates toxic leaves
by adding one extra query to retrieve the value
of the desired symbol
(this is also why the RLDC must be systematic, so that $\msg_i$ is part of $\Code(\msg)$).
This is the query that we wish to save in \cref{thm:stronger-goldberg}. To avoid losing this extra query, we will isolate the toxic leaves
by first relabeling all non-toxic leaves,
so that all of the remaining completeness error
is caused only by toxic leaves. 
Then, the following lemma shows that toxic leaves are rarely chosen:
\begin{lemma}
  \label{lem:toxic-rare}
  Let $\Dec_R(i)$
  be the rerandomized decoder for index $i$
  after all non-toxic leaves have been relabeled.
  Suppose it has completeness error $\epsilon$
  which is caused only by toxic leaves. Then,
  for all $\msg \in \F^k$,
  \begin{equation*}
    \Pr[\text{\textnormal{$\Dec_R^{\Code(\msg)}(i)$ ends on a toxic leaf}}]
    \le \frac{\abs{\F}}{\abs{\F} - 1} \cdot \epsilon \mper
  \end{equation*}
\end{lemma}
\begin{proof}
  Intuitively, the best possible behavior on a toxic leaf
  is to guess a random symbol from $\F$,
  in order to minimize the worst-case completeness error
  across all codewords.
  The probability of guessing correctly is $1/\abs{\F}$,
  and so we should get a $\abs{\F}/(\abs{\F}- 1)$ factor
  relative to $\epsilon$.

  Formally, pick $\abs{\F}$ messages $\msg^1, \dots, \msg^{\abs{\F}}$ where $\msg^j$ has $i$-th symbol $j \in \F$.
  All of these codewords have the same completeness error
  because $\Dec_R(i)$ rerandomizes the input. We then have
  \begin{align*}
    \epsilon = \Pr[\Dec_R^{\Code(\msg)}(i) \neq \msg_i]
    =
    \sum_{\text{toxic leaves $\ell$}}
    \Pr[\text{leaf $\ell$ chosen}]
    \cdot\Pr[\Dec_R^{\Code(\msg)}(i) \neq \msg_i
    \mid \text{leaf $\ell$ chosen}]
  \end{align*}
  Because $\Dec_R$ rerandomizes over the input and all of the
  $\msg^j$ are codewords,
  the probability of picking each particular leaf
  is the same for all of the different messages. Hence,
  \begin{align*}
    \abs{\F} \epsilon
    &= \sum_j \Pr[\Dec_R^{\Code(\msg^j)}(i) \neq \msg^j_i]\\
    &=
    \sum_{\text{toxic leaves $\ell$}}
    \Pr[\text{leaf $\ell$ chosen}]
    \cdot\left( {
        \sum_j \Pr[\Dec_R^{\Code(\msg^j)}(i) \neq \msg^j_i \mid \text{leaf
      $\ell$ chosen}]}
    \right)\\
    &=
    \sum_{\text{toxic leaves $\ell$}}
      \Pr[\text{leaf $\ell$ chosen}] \cdot
      \#\{\alpha \in \F : \text{$\ell$ is \emph{not} labeled with $\alpha$} \}\\
    &\ge (\abs{\F} - 1) \cdot \Pr[\text{toxic leaf chosen}]\mper\qedhere
  \end{align*}
\end{proof}

We now know how often toxic leaves are selected by the decoder,
which will help us remove them from the query distribution later on.
For now, however, we can relabel them to make additional queries
and achieve perfect completeness.
When the decoder ends up on a toxic leaf,
run a \emph{global decoding} subroutine which queries the entire input
and decodes the $i$-th message index;
this shifts the remaining completeness error to soundness error
using \cref{lem:leaf-relabeling}.
We use global decoding to avoid requiring that the RLDC is systematic,
which was necessary in the original proof of Goldberg.
In summary, we have shown the following transformation:
\begin{proposition}
  \label{lem:relabel-summary}
  Let $\Code\colon \F^k \to \F^n$ be a linear $(q, \delta, 1-\epsilon, s)$-RLDC (or
  RLCC).
  Then, $\Code$ has an adaptive relaxed local decoder (or corrector)
  with perfect completeness and
  soundness error at most $s + \epsilon$.
  This decoder, when given a valid codeword as input,
  selects a non-toxic leaf (and makes $q$ queries) with probability $\ge 1 - (\abs{\F}
  \epsilon)/(\abs{\F} - 1)$,
  and selects a toxic leaf (and makes $n$ queries) otherwise.
\end{proposition}

\subsection{Removing adaptivity and pruning toxic leaves}
Lastly, Goldberg removes adaptivity by showing that
we can evaluate the decision tree distribution
on a uniformly random codeword to determine the index set to query.
Now that the queries are nonadaptive, we can
condition on not picking a toxic leaf,
which adds a bit more soundness error.
\begin{lemma}[{\cite[Lemma 20]{Goldberg24}}]
  \label{lem:nonadaptive-step}
  If $\Code$ has an adaptive relaxed local decoder $\Dec$
  with perfect completeness and soundness error $s$,
  then $\Code$ has a \emph{nonadaptive}
  and \emph{canonical} relaxed local decoder
  with the same completeness and soundness,
  by selecting a uniformly random codeword $c$
  and
  simulating $\Dec$ on $c$ to determine its query set $Q$.
\end{lemma}
\begin{proof}[Proof of \cref{thm:stronger-goldberg}]
  Using \cref{lem:relabel-summary},
  we can start with $\Code$ which is a $(q, \delta, 1-\epsilon, s)$-RLDC
  and get a relaxed local decoder with perfect completeness
  and soundness error $s + \epsilon$,
  and which selects a relabeled toxic leaf
  with probability at most
  $p = (\abs{\F}
  \epsilon)/(\abs{\F} - 1)$.
  Then, use \cref{lem:nonadaptive-step}
  to get a nonadaptive decoder
  with the same completeness and soundness.
  Finally, modify the query distribution to
  condition on never selecting a toxic leaf;
  the perfect completeness is unharmed
  but the soundness error will increase from $s + \epsilon$
  to $(s + \epsilon)/(1-p) \leq s + \eps + p$, which is $s + \epsilon(2\abs{\F} - 1)/(\abs{\F} - 1)$.
  Now, we are also guaranteed that the decoder
  makes $q$ queries, which finishes the proof.
\end{proof}

\section{Constructions of Small Query RLDCs/RLCCs That Are Not LDCs}
\label{sec:constructions}
In this section, we give a simple family of explicit codes that are $q$-RLDCs/RLCCs where $q$ is a small, explicit constant, and are not $q$-LDCs for any constant $q$.
\begin{theorem}[Formal \cref{mthm:rldcconstructions}]
\label{thm:formalrldcconstruction}
There is a linear code $\Code \colon \F_2^{k} \to \F_2^{N}$ where $N = k^{O(\log \log k)}$ such that for some constant $\delta > 0$, $\Code$ is:
\begin{inparaenum}[(1)]
\item a $(15, \delta, 1, 1/3)$-RLDC;
\item a $(41, \delta, 1, 1/2 - \eps)$-RLCC for a small constant $\eps > 0$;
\item a $(58, \delta, 1, 1/3)$-RLCC;
\item not a $(O(\log k), O(N^{-1/3}), 1, 1/2 - \eps)$-LDC for any (including subconstant) $\eps > 0$.
\end{inparaenum}
\end{theorem}
We will prove \cref{thm:formalrldcconstruction} in the next $3$ subsections. In~\cref{sec:construction}, we will define our code. Then, in~\cref{sec:rldcconstruction}, we will prove that it is an RLDC, and in~\cref{sec:rlccconstruction}, we will prove that is an RLCC. Finally, in \cref{sec:notldc} we will prove that it is not an LDC.

\subsection{The construction of the code}
\label{sec:construction}
In this section, we will define the code $\Code$ in \cref{thm:formalrldcconstruction}. Let $t \in \N$ and let $\F_{2^t}$ be the finite field with $2^t$ elements. We recall the following basic facts about finite fields.
\begin{fact}[Finite field notation]
\label{fact:finitefields}
Let $\F_{2^t}$ be the finite field with $2^t$ elements. The field $\F_{2^t}$ is an $\F_2$-vector space of dimension $t$, and therefore there is an $\F_2$-linear isomorphism $\pi \colon \F_{2^t} \to \F_2^t$.  The map $\pi$ depends on the choice of basis for $\F_{2^t}$, which we will view as fixed in advance.

For any $\alpha \in \F_{2^t}$ and $i \in [t]$, we let $\pi(\alpha)_i$ denote the $i$-th coordinate of $\alpha$. Because multiplication by $\alpha$ is an invertible $\F_2$-linear transformation in $\F_{2^t}$, there exists a $t \times t$ matrix $M_{\alpha} \in \F_2^{t \times t}$ such that for any $\beta \in \F_{2^t}$, $\pi^{-1}(M_{\alpha} \pi(\beta)) = \alpha \beta$. In particular, for each $\alpha \in \F_{2^t}$ and $i \in [t]$, there exists $v \in \F_2^t$ such that for any $\beta \in \F_{2^t}$, $\ip{v, \pi(\beta)} = \pi(\alpha \beta)_i$.
\end{fact}

The code $\Code$ is defined formally via its encoding map. We will first define the linear subspace of codewords, and then explain how to define the encoding map. 

Let $n, d \in \N$ be parameters with $d < 2^t$, and let $k = {n \choose \leq d} \defeq \sum_{i = 0}^d {n \choose i}$. Let $f \colon \F_{2^t}^n \to \F_{2^t}$ be a polynomial of degree at most $d$ in $n$ variables $x_1, \dots, x_n$. For each line $L$ in $\F_{2^t}^n$, let $S_L$ be an arbitrary (ordered) subset of $L$ of size $d+1$. For a collection of field elements $(\alpha_1, \dots, \alpha_{d+1}) \in \F_{2^t}^{d+1}$, we let $\Had(\alpha_1, \dots, \alpha_{d+1})$ be the encoding of these field elements using the Hadamard code over $\F_2$ and the map $\pi$. That is, $\Had(\alpha_1, \dots, \alpha_{d+1})$ is a vector of length $2^{t(d+1)}$, where entries are indexed by $v = (v_1, \dots, v_{d+1}) \in \F_2^{t(d+1)}$, and the $v$-th entry is $\sum_{i = 1}^{d+1} \ip{v_i,\pi(\alpha_i)}$. 

With the above setup, we can now specify the set of codewords. For each polynomial $f \colon \F_{2^t}^n \to \F_{2^t}$ of degree at most $d$, we obtain a codeword by encoding the function $f$ via the bit-wise concatenation of $\Had(f(S_L))$ for each line $L$. That is, the codeword corresponding to $f$ has, for each line $L$, a block of $2^{t(d+1)}$ bits that is $\Had(f(S_L))$, so that we can view the codeword as a collection $\{\Had(f(S_L))\}_L$ indexed by all lines $L$ in $\F_{2^t}^n$.

\parhead{Setting parameters.} The set of codewords is clearly an $\F_2$-linear subspace, and it has dimension $tk$, where $k = {n \choose \leq d}$. Moreover, it is a linear subspace in $\F_{2}^{\#L \cdot 2^{t(d+1)}}$, where $\#L = 2^{tn} (2^{tn} - 1)/(2^t(2^t - 1))$ is the number of lines in $\F_{2^t}^n$. To minimize the blocklength as a function of $k$, we take $d = O(n)$ (which forces $t = \Theta(\log d) = \Theta(\log n)$, as we need $2^t > d$), so that $k = 2^{O(n)}$ and the blocklength is $2^{O(n \log n)}$. Hence, the blocklength is $k^{O(\log \log k)}$.

\parhead{Choosing an encoding map.} To finish defining the code $\Code$, we need to specify the encoding map. To do this, we first choose an arbitrary $\F_{2^t}$-linear map $A$ from $\F_{2^t}^k$ to the set of degree $\leq d$ polynomials $f$ with the property that there are elements $x^{(1)}, \dots, x^{(k)} \in \F_{2^t}$ such that if $f = A b$ for $b \in \F_{2^t}$, then $f(x^{(i)}) = b_i$ for all $i$. We can then extend $A$ to a map from $\F_{2}^{tk}$ to the set of degree $\leq d$ polynomials by splitting the input into $k$ blocks of size $t$, applying the linear map $\pi^{-1}$ (\cref{fact:finitefields}) on each block, applying $A$ to the output, and then applying $\pi$ to each field element in the evaluation table of the resulting polynomial $f$.

As we will show, our RLDC decoder, when given access to a corrupted version of the codeword corresponding to the polynomial $f$, will be able to recover $\pi(\alpha f(x))_i$ for every $\alpha \in \F_{2^t}$, $x \in \F_{2^t}^n$, and any $i \in [t]$. As a result, the specific choice of encoding map $A$ specified above is not important.

\parhead{Comparison to~\cite{AsadiS21}.}
Our construction shares some similarities to the work of~\cite{AsadiS21}. Similar to our construction, they first use a Reed--Muller code to encode a message as a polynomial $f$, and then they encode each ``local view'' of $f$ in some way. Their ``local views'', however, are a special set of planes $\cP$ that have ``directions'' in $\mathbb{H}^n$, where $\mathbb{H}$ is a subfield of $\F$. They then encode $f \vert_{\cP}$ for each plane $\cP$ using a canonical correctable Probabilistically Checkable Proof of Proximity (ccPCPP). Their decoder then takes certain random walk of length $[\mathbb{F} : \mathbb{H}] + 1$ on the special planes $\cP$, and uses the ccPCPPs to decode values of $f$ and do consistency checks.

In contrast, we encode $f$ by encoding its ``local view'' $f \vert_L$ for every line $L$ using the Hadamard code. As we shall see, our RLDC decoder takes a random walk of length $2$ on the lines, and our RLCC decoder takes a random walk of length $3$.

\parhead{An observation about $\Code$.} Finally, we make some observations about $\Code$, which will be useful in the proofs.
\begin{observation}
\label{obs:codeobs}
Fix a line $L$. For any $z \in L$, any $\alpha \in \F_{2^t}$, and any $i \in [t]$, there exists a point $v^{L,z, \alpha, i} \in \F_2^{t(d+1)}$ such that $\Had(f(S_L))(v^{L,z, \alpha, i}) = \pi(\alpha f(z))_i$. That is, for any $i \in [t]$, one can recover the $i$-th bit of $\alpha \cdot f(z)$ using the Hadamard encoding of (only) $d+1$ points $S_L$ on the line $L$.
\end{observation}
\medskip
\noindent The point $v^{L,z,\alpha,i}$ should be interpreted as: to recover the $i$-th bit of $\alpha f(z)$ using the Hadamard encoding of $f \vert_L$, we query the Hadamard encoding of $f \vert_L$ at the point $v^{L, z ,\alpha, i}$.
\begin{proof}
Let $z_1, \dots, z_{d+1}$ be the points in $S_L$. Because $f$ is a degree $d$ polynomial and $d < 2^t = \abs{\F_{2^t}}$, by polynomial interpolation, there exist coefficients $\alpha_1, \dots, \alpha_{d+1} \in \F_{2^t}$ such that $f(z) = \sum_{j = 1}^{d+1} \alpha_j f(z_j)$. Therefore, $\pi(\alpha f(z))_i = \sum_{j = 1}^{d+1} \pi(\alpha \alpha_j f(z_j))_i$. For each $j$, let $v^{j}$ be the $i$-th row of $M_{\alpha \alpha_j}$, the matrix defined in~\cref{fact:finitefields}. The vector $v^j$ then has the property that $\ip{v^j, \pi(\beta)} = \pi(\alpha \alpha_j \beta)_i$ for all $\beta \in \F_{2^t}$, and therefore $\pi(\alpha \alpha_j f(z_j))_i = \ip{v^j, \pi(f(z_j))}$. Concatenating the vectors $v^j$ together yields the vector $v^{L,z, \alpha, i}$.
\end{proof}

\subsection{The RLDC decoder and its analysis}
\label{sec:rldcconstruction}
In this subsection, we prove Item (1) in \cref{thm:formalrldcconstruction}, where the code is defined in \cref{sec:construction}. 

\begin{proof}[Proof of Item (1) in \cref{thm:formalrldcconstruction}]
To show that the code is an RLDC, we will analyze the decoder defined below.
\begin{mdframed}
  \begin{algorithm}
    \label{alg:rldc}\mbox{}
    \begin{description}
    \item[Given:]
       A  collection of functions $\{G_{L}\}_{L}$ where each $G_{L} \colon \F_{2}^{t(d+1)} \to \F_{2}$ is an arbitrary function, along with a point $x^* \in \F_{2^t}^n$, a field element $\alpha^* \in \F_{2^t}$, and an index $i^* \in [t]$. The collection $\{G_{L}\}_{L}$ is supposed to be equal to $\{\Had(f(S_L))\}_{L}$ for some polynomial $f$ of degree $\leq d$.
    \item[Output:]
        A symbol in $\{0,1,\bot\}$, hopefully equal to $\pi(\alpha^* f(x^*))_{i^*}$.
           \item[Operation:]\mbox{}

    \begin{enumerate}[(1)]
	\item Choose a line $L^*$ containing $x$ uniformly at random.
	\item \textbf{Run $r_1$ linearity tests on $G_{L^*}$:} pick random $v^{(1)}, v^{(2)}, v^{(3)} \in \F_{2}^{t(d+1)}$, and check that $G_{L^*}(v^{(1)}) + G_{L^*}(v^{(2)}) + G_{L^*}(v^{(3)}) = 0$. If the check fails, output $\bot$. Repeat $r_1$ times.
	\item \textbf{Run $r_2$ consistency tests:}
	\begin{enumerate}[(a)]
	\item Pick $y \in L^*$ uniformly at random, $\alpha \in \F_{2^t}$ uniformly at random, and $i \in [t]$.
	\item \textbf{Locally decode $\pi(\alpha f(y))_i$ from $G_{L^*}$:} let $v^{L,y,\alpha,i}$ be the vector from \cref{obs:codeobs}. Let $v^{(4)} \in \F_{2}^{t(d+1)}$ be chosen uniformly at random. Let $a_{L^*} = G_{L^*}(v^{L^*,y,\alpha,i} + v^{(4)}) + G_{L^*}(v^{(4)})$.
	\item \textbf{Locally decode $\pi(\alpha f(y))_i$ from $G_{L'}$:} choose $L'$ to be a uniformly random line containing $y$. Let $v^{L',y,\alpha,i}$ be the vector from \cref{obs:codeobs}. Let $v^{(5)} \in \F_{2}^{t(d+1)}$ be chosen uniformly at random. Let $a_{L'} = G_{L'}(v^{L',y,\alpha,i} + v^{(5)}) + G_{L'}(v^{(5)})$.
	\item \textbf{Consistency check:} check that $a_{L^*} = a_{L'}$ and output $\bot$ if the check fails.
	\item Repeat $r_2$ times.
	\end{enumerate}
			\item \textbf{Locally decode $\pi(\alpha^*f(x^*))_{i^*}$ from $G_{L^*}$:} let $v^{L^*,x^*,\alpha^*,i^*}$ be the vector from \cref{obs:codeobs}. Let $v^{(6)} \in \F_{2}^{t(d+1)}$ be chosen uniformly at random. Output $G_{L^*}(v^{L^*,x^*,\alpha^*,i^*} + v^{(6)}) +G_{L^*}(v^{(6)})$.
      \end{enumerate}
    \end{description}
  \end{algorithm}
\end{mdframed}
In the above algorithm, $r_1$ and $r_2$ are positive integers, which we will choose later.

First, we observe that by \cref{obs:codeobs}, \cref{alg:rldc} returns $\pi(\alpha^* f(x^*))_{i^*}$ for any input $(x^*, \alpha^*, i^*)$ if the collection of functions is indeed $\{\Had(f(S_L))\}_{L}$ for some polynomial $f$ of degree $\leq d$. We note that our choice of encoding map, the message bits correspond to evaluations of $f$ on specific points, and so this allows us to recover any bit of the message. Thus, \cref{alg:rldc} has perfect completeness. We also note that \cref{alg:rldc} makes $3r_1 + 4r_2 + 2$ queries, though we will explain how to slightly reduce the query complexity later.

We now analyze the soundness error of \cref{alg:rldc}. For each line $L$, let $H_L$ be the closest linear function to $G_L$. Given a linear function $H_L$, there is a unique univariate degree $d$ polynomial $h_L$ on the line $L$ such that for each $z \in L$, $\alpha \in \F_{2^t}$, and $i \in [t]$, $\pi(\alpha h_L(z))_i = H_L(v^{L, z, \alpha, i})$. The polynomial $h_L$ is simply defined by ``extracting'' the values of $h_L$ on $S_L$ using \cref{obs:codeobs} and then defining $h_L$ on the rest of the line using polynomial interpolation. We let $F_L$ be the linear function $\Had(f(S_L))$, and we define $f_L$ to be the polynomial $f \vert_L$. We note that the process used to obtain $h_L$ from $H_L$ yields $f_L$ if $H_L = F_L$. For two Boolean functions $F_L$ and $G_L$, we let $\bar{\Delta}(F_L, G_L)$ denote the (relative) Hamming distance over $\F_2$, and for two polynomials $f_L$ and $g_L$, we let $\bar{\Delta}(f_L, g_L)$ denote the (relative) Hamming distance over $\F_{2^t}$, i.e., the fraction of $x \in L$ such that $f_L(x) \ne g_L(x)$.

We will split our analysis into the following disjoint cases, and bound the probability that decoder errs in each case.
\begin{enumerate}[(1)]
\item The random line $L^*$ through $x^*$ is ``bad'', in that $\E_{L' : \abs{L' \cap L^*} \geq 1, x^* \notin L'}[\bar{\Delta}(G_{L'}, F_{L'})] \geq \delta^{2/3}$.
\item $f_{L^*} \ne h_{L^*}$ and $\E_{L' : \abs{L' \cap L^*} \geq 1, x^* \notin L'}[\bar{\Delta}(G_{L'}, F_{L'})] \leq \delta^{2/3}$, but both the repeated linearity test and the repeated consistency test pass.
\item $f_{L^*} = h_{L^*}$, but the repeated linearity test passes and the output is $1 - \pi(\alpha^* f(x^*))_{i^*}$ (i.e., not in $\{\pi(\alpha^* f(x^*))_{i^*}, \bot\}$).
\end{enumerate}

\parhead{Analysis of Case 1.} Recall that the collection $\{G_L\}_{L}$ is $\delta$-close to $\{F_L\}_{L}$. Equivalently, this means that $\E_{L}[\bar{\Delta}(G_L, F_L)] \leq \delta$. Since $L^*$ is a uniformly random line passing through $x^*$, and $L'$ is a uniformly random line passing through a random point $y \ne x^*$ on $L^*$, it follows that $L'$ is distributed as a uniformly random line. Therefore, $\E_{L} [\E_{L' : \abs{L' \cap L} \geq 1, x^* \notin L'}[\bar{\Delta}_L(G_L, F_L)]] = \delta$. It follows by Markov's inequality that the probability over $L$ that $\E_{L' : \abs{L' \cap L} \geq 1, x^* \notin L'}[\bar{\Delta}_L(G_L, F_L)] \geq \delta^{2/3}$ is at most $\delta^{1/3}$.

\parhead{Analysis of Case 2.} Let $\eps= \bar{\Delta}(G_{L^*}, H_{L^*})$. Recall by \cref{fact:linearitytest}, the probability that one iteration of the linearity test passes is at most $1 - \bar{\Delta}(G_{L^*}, H_{L^*}) = 1 - \eps$. Hence, the probability that all $r_1$ repetitions pass is at most $(1 - \eps)^{r_1}$.

Let us now analyze one iteration of the consistency test. Call the line $L'$ chosen ``bad'' if $\bar{\Delta}(G_{L'}, F_{L'}) \geq \delta^{1/3}$. Since $\E_{L' : \abs{L' \cap L^*} \ge 1, x^* \notin L'}[\bar{\Delta}(G_{L'}, F_{L'})] \leq \delta^{2/3}$, it follows that the probability that $L'$ is ``bad'' is at most $\delta^{1/3}$.

Let us proceed assuming that $L'$ is not ``bad''. Then, since $F_{L'}$ is a linear function, by \cref{fact:linrecovery} and \cref{obs:codeobs}, it follows that $a_{L'}$ is equal to $\pi(\alpha f_{L^*}(y))_i$ (which is $\pi(\alpha f(y))_i$) with probability at least $1 - 2 \delta^{1/3}$. Similarly, because $\bar{\Delta}(G_{L^*}, H_{L^*}) = \eps$, it follows that $a_{L^*}$ is equal to $\pi(\alpha h_{L^*}(y))_i$ with probability at least $1 - 2\eps$. Moreover, this holds regardless of the choice of $\alpha$ and $i$.

Finally, because $f_{L^*} \ne h_{L^*}$ and these are different degree $d$ polynomials, the probability that $f_{L^*}(y) \ne h_{L^*}(y)$ is at least $1 - d/n$. And, if such a $y$ is chosen, the probability that $\pi(\alpha f_{L^*}(y))_i \ne \pi(\alpha h_{L^*}(y))_i$ is $1/2$, as $\alpha(f_{L^*}(y) - h_{L^*}(y))$ is a random element of $\F_{2^t}$, and so its $i$-th bit is $1$ with probability $1/2$.

Thus, we can conclude that each round of the consistency test passes with probability at most $\delta^{1/3} + d/n + (1 - d/n)(\frac{1}{2} + \frac{1}{2} (2 \delta^{1/3} + 2 \eps))$. Since the tests are independent, we conclude that the probability that all rounds of both tests pass is at most $(1 - \eps)^{r_1} \left(\delta^{1/3} + d/n + (1 - d/n)(\frac{1}{2} + \delta^{1/3} + \eps)\right)^{r_2}$.

\parhead{Analysis of Case 3.} By the previous analysis, the probability that all linearity tests pass is at most $(1 - \eps)^{r_1}$, where $\eps= \bar{\Delta}(G_{L^*}, H_{L^*})$. By \cref{fact:linrecovery}, the output is $\pi(h_{L^*}(x^*))_{i^*}$ with probability at least $1 - 2 \eps$, which is $\pi(f_{L^*}(x^*))_{i^*} = \pi(f(x^*))_{i^*}$ since $h_{L^*} = f_{L^*}$. Thus, the probability that the decoder outputs an incorrect answer is at most $(1 - \eps)^{r_1} \cdot2 \eps$ in this case.

\bigskip
In total, we conclude that the probability that the output of the decoder is not in $\{\pi(f(x^*))_i, \bot\}$ is at most
\begin{equation*}
\eta(\eps) = \max\{\delta^{1/3}, (1 - \eps)^{r_1}\left(\delta^{1/3} + d/n + (1 - d/n)(1/2 +  \delta^{1/3} +  \eps)\right)^{r_2}, (1 - \eps)^{r_1} \cdot 2 \eps\} \mcom
\end{equation*}
where $\eps =  \bar{\Delta}(G_{L^*}, H_{L^*})$. Setting $r_1 = r_2 = 2$, $\delta$ and $d/n$ to be sufficiently small constants, and taking the maximum over all $\eps$ shows that $\eta \leq \frac{1}{2} - \eps'$ for some constant $\eps'$. In fact, by taking $\delta$ and $d/n$ to be sufficiently small constants, we can make $\eta = (3/4)^4 + \eps'' \leq 1/3$ for a small constant $\eps''$.

The query complexity is $3r_1 + 4r_2 + 2 = 16$. Below, we shall explain how to save one query.

\parhead{Saving a query.} Observe that in the above analysis, we analyzed Cases (2) and (3) separately. This allows us to ``recycle'' our queries across the different cases. Namely, we observe that in Item (3b) in \cref{alg:rldc}, it holds that $v^{(4)}$ is chosen uniformly at random. Because we do not use the consistency test to analyze Case (3) above, we can reuse this query by taking $v^{(6)} = v^{(4)}$ and the analysis does not change. This allows us to make the query complexity slightly smaller: $3r_1 + 4r_2 + 1 = 15$.
\end{proof}

\subsection{The RLCC decoder and its analysis}
\label{sec:rlccconstruction}
In this subsection, we prove Items (2) and (3) in \cref{thm:formalrldcconstruction}. 
The key difference between the RLCC decoder and RLDC decoder is that the RLCC decoder may be asked to decode an arbitrary point $v \in \F_2^{t(d+1)}$ on the Hadamard encoding of some line $L^*$. Unlike in the case of the RLDC decoder, where the point $v$ corresponded to decoding $\pi(f(x))_i$ for some point $x$ and some index $i$, here the point $v$ might not correspond to any evaluation point of the polynomial $f$.

\begin{proof}[Proof of Items (2) and (3) in~\cref{thm:formalrldcconstruction}]
To show that the code is an RLCC, we will analyze the decoder defined below.
\begin{mdframed}
  \begin{algorithm}
    \label{alg:rlcc}\mbox{}
    \begin{description}
    \item[Given:]
       A  collection of functions $\{G_{L}\}_{L}$ where each $G_{L} \colon \F_{2}^{t(d+1)} \to \F_{2}$ is an arbitrary function, along with a point $v^* \in \F_2^{t(d+1)}$ and a line $L^*$. The collection $\{G_{L}\}_{L}$ is supposed to be equal to $\{\Had(f(S_L))\}_{L}$ for some polynomial $f$ of degree $\leq d$.
    \item[Output:]
        A symbol in $\{0,1,\bot\}$, hopefully equal to $\Had(f(S_{L^*}))(v^*)$.
           \item[Operation:]\mbox{}

    \begin{enumerate}[(1)]
	\item \textbf{Run $r_1$ linearity tests on $G_{L^*}$:} pick random $v^{(1)}, v^{(2)}, v^{(3)} \in \F_{2}^{t(d+1)}$, and check that $G_{L^*}(v^{(1)}) + G_{L^*}(v^{(2)}) + G_{L^*}(v^{(3)}) = 0$. If the check fails, output $\bot$. Repeat $r_1$ times.
	\item \textbf{Run $r_2$ consistency tests times:}
	\begin{enumerate}[(1)]
	\item Pick $y \in L^*$ uniformly at random, $\alpha \in \F_{2^t}$ uniformly at random, and $i \in [t]$.
	\item \textbf{Locally decode $\pi(\alpha f(y))_i$ from $G_{L^*}$:} let $v^{L,y,\alpha,i}$ be the vector from \cref{obs:codeobs}. Let $v^{(4)} \in \F_{2}^{t(d+1)}$ be chosen uniformly at random. Let $a_{L^*} = G_{L^*}(v^{L^*,y,\alpha,i} + v^{(4)}) + G_{L^*}(v^{(4)})$.
	\item \textbf{Locally decode $\pi(\alpha f(y))_i$ using the RLDC decoder:} Run the RLDC decoder in \cref{alg:rldc} to recover $\pi(\alpha f(y))_i$. If the RLDC decoder outputs $\bot$, then abort and output $\bot$. Else, let $a_{\mathrm{RLDC}} \in \Bits$ be the output of the RLDC decoder.
	\item \textbf{Consistency check:} check that $a_{L^*} = a_{\mathrm{RLDC}}$ and output $\bot$ if the check fails.
	\item Repeat $r_2$ times.
	\end{enumerate}
			\item \textbf{Locally decode $\Had(f(S_{L^*}))(v^*)$ from $G_{L^*}$:} Let $v^{(6)} \in \F_{2}^{t(d+1)}$ be chosen uniformly at random. Output $G_{L^*}(v^* + v^{(6)}) +G_{L^*}(v^{(6)})$.
      \end{enumerate}
    \end{description}
  \end{algorithm}
\end{mdframed}
The analysis of \cref{alg:rlcc} is similar to the analysis of \cref{alg:rldc}. Below, we will reuse the same notation as done in \cref{sec:rldcconstruction}.
The fact that the decoder has perfect completeness is straightforward, and so we proceed with analyzing the soundness error.

As before, will split our analysis into disjoint cases, and bound the probability that decoder errs in each case.
\begin{enumerate}[(1)]
\item $f_{L^*} \ne h_{L^*}$, but both the repeated linearity test and repeated consistency test pass.
\item $f_{L^*} = h_{L^*}$, but the repeated linearity test passes and the output is $1 - \Had(f(S_{L^*}))(v^*)$ (i.e., not in $\{\Had(f(S_{L^*}))(v^*), \bot\}$).
\end{enumerate}

\parhead{Analysis of Case 1.}
The probability that all $r_1$ repetitions of the linearity pass when $\bar{\Delta}(G_{L^*}, H_{L^*}) =\eps$ is at most $(1 - \eps)^{r_1}$ by \cref{fact:linearitytest}.

Let us now analyze one iteration of the consistency test. Because $f_{L^*} \ne h_{L^*}$ and these are different degree $d$ polynomials, the probability that $f_{L^*}(y) \ne h_{L^*}(y)$ is at least $1 - d/n$. Suppose that such a $y$ is chosen. Then, the probability that $\pi(\alpha f_{L^*}(y))_i \ne \pi(\alpha h_{L^*}(y))_i$ is $1/2$, as $\alpha(f_{L^*}(y) - h_{L^*}(y))$ is a random element of $\F_{2^t}$, and so its $i$-th bit is $1$ with probability $1/2$.

Let us assume that $y$, $\alpha$ and $i$ are ``good'', meaning that $\pi(\alpha f_{L^*}(y))_i \ne \pi(\alpha h_{L^*}(y))_i$. Because $\bar{\Delta}(G_{L^*}, H_{L^*}) \leq \eps$, it follows that $a_{L^*}$ is equal to $\pi(\alpha h_{L^*}(y))_i$ with probability at least $1 - 2\eps$. Moreover, this holds regardless of the choice of $\alpha$ and $i$.

We now invoke the soundness of the RLDC decoder. The RLDC decoder either outputs the correct bit $\pi(\alpha f_{L^*}(y))_i$, or else it outputs $\bot$, with probability at least $1 - \eta_{\mathrm{RLDC}}$ for some constant $\eta_{\mathrm{RLDC}}$. Hence, the probability that the RLDC decoder outputs the wrong bit is at most $\eta_{\mathrm{RLDC}}$.

Thus, the probability that one round of the consistency test passes is at most $d/n + \frac{1}{2}(1-d/n) + \frac{1}{2}(1 - d/n)(\eta_{\mathrm{RLDC}} + (1 - \eta_{\mathrm{RLDC}})2\eps)$. Because the consistency test and linearity tests are independent, the total probability of all tests passing is at most

\begin{equation*}
(1 - \eps)^{r_1} \left(d/n + \frac{1}{2}(1 - d/n)(1 + \eta_{\mathrm{RLDC}} + (1 - \eta_{\mathrm{RLDC}})2\eps)\right)^{r_2} \mper
\end{equation*}

\parhead{Analysis of Case 2.} The probability that all $r_1$ repetitions of the linearity pass when $\bar{\Delta}(G_{L^*}, H_{L^*}) =\eps$ is at most $(1 - \eps)^{r_1}$ by \cref{fact:linearitytest}. By \cref{fact:linrecovery}, the output is $h_{L^*}(v^*)$ with probability at least $1 - 2 \eps$, which is $\Had(f(S_{L^*}))(v^*)$ since $h_{L^*} = f_{L^*}$. As these are independent, the probability that the decoder outputs an incorrect answer is at most $(1 - \eps)^{r_1} \cdot 2 \eps$ in this case.

\bigskip

In total, we conclude that the probability that output of the decoder is incorrect is at most
\begin{equation*}
\eta(\eps) = \max\{(1 - \eps)^{r_1} \left(d/n + \frac{1}{2}(1 - d/n)(1 + \eta_{\mathrm{RLDC}} + (1 - \eta_{\mathrm{RLDC}})2\eps)\right)^{r_2}, (1 - \eps)^{r_1} \cdot 2 \eps\} \mcom
\end{equation*}
where $\eps = \bar{\Delta}(G_{L^*}, H_{L^*})$. Recall that $\eta_{\mathrm{RLDC}} < 1/3$ provided that $\delta$ is a sufficiently small constant. Hence, setting $r_1 = r_2 = 2$, $d/n$ to be a sufficiently small constant, and maximizing $\eta(\eps)$ over all $\eps$, we see that the maximum is at most $\frac{4}{9} + \eps'' < 1/2$ for some sufficiently small constant $\eps''$.

Setting $r_1 = 2$ and $r_2 = 3$, we get that $\eta \leq (2/3)^3 + \eps'' < 1/3$ for a sufficiently small constant $\eps''$.

The query complexity of the decoder is $3r_1 + (2 + q_{\mathrm{RLDC}})r_2 + 1$ (using the trick to reduce the query complexity by $1$ from~\cref{sec:rldcconstruction}). Since $q_{\mathrm{RLDC}} = 15$, this yields $41$ queries for the case of soundness error below $1/2$, and $58$ queries for the case of soundness error below $1/3$.
\end{proof}
\subsection{The code is not an LDC}
\label{sec:notldc}
In this section, we prove Item (4) in~\cref{thm:formalrldcconstruction}, i.e., that the code constructed in \cref{sec:construction} is not an LDC. 
\begin{proof}[Proof of Item (4) in~\cref{thm:formalrldcconstruction}]
Recall that the code is defined as follows. For each degree $\leq d$ polynomial $f \colon \F_{2^t}^n \to \F_{2^t}$ in $n$ variables, we encode $f$ by writing down $\Had(f(S_L))$ for each line $L$ in $\F_{2^t}^n$, where we have set $d = \Theta(n)$, $t = \Theta(\log d) = \Theta(\log n)$, and we have $k = t {n \choose \leq d}$. The blocklength of the code is $N = \#L \cdot 2^{t(d+1)}$, where $\#L = 2^{tn} (2^{tn} - 1)/(2^t(2^t - 1))$ is the number of lines in $\F_{2^t}^n$. We will show that $\Code$ is not a $(q, \delta, 1, 1/2 - \eps)$-LDC for any $\eps > 0$ (that need not be constant) for $q = d$ and $\delta = 2^{-n(t-1)}$. Note that $q = d \leq O(\log k)$ and $\delta = 2^{-n(t-1)} \leq O(N^{-1/3})$.

\parhead{Setup.} 
To show that the code is not a $(d, 2^{-n(t-1)}, 1, 1/2 - \eps)$-LDC for any $\eps > 0$, it suffices to show that for any $d$-query LDC decoder $\Dec$  and any choice of $\alpha^* \in \F_{2^t}$, $x^* \in \F_{2^t}^n$, and $i^* \in [t]$, there is a collection of ``local encodings'' $\{h_L\}_{L}$ that is $2^{-n(t-1)}$-close to a codeword $\{\Had(f(S_L))\}_{L}$ such that $\Pr[\Dec^{\{h_L\}_{L}}(\alpha^*, x^*, i^*)] = \frac{1}{2}$.

First, we observe that it suffices to argue this in the \emph{erasure} error model, as opposed to the standard Hamming error model. In the erasure model, we corrupt a codeword by replacing a bit in the encoding with an error symbol $\bot$ to signify that the bit has been erased. That is, the outputs of the ``local encodings'' $h_L$ can only agree with $\Had(f(S_L))$ or be $\bot$. Clearly, if a decoder works in the standard Hamming error model, then it also works in the erasure error model, as we can simulate the Hamming error model by replacing any $\bot$ symbol by a $0$. Hence, it suffices to argue that there is no decoder in the erasure error model. In fact, it suffices to argue this even in the following weaker ``line erasure error model'', where we are only allowed to erase entire lines. That is, for each line $L$, we either have $h_L \equiv \Had(f(S_L))$, i.e., they are the same function, or $h_L \equiv \bot$, i.e., the entire function outputs $\bot$.

Now that we are working in the line erasure error model, we will strengthen the decoder by allowing it to make ``line queries''. That is, instead of 
allowing the decoder to make one query to retrieve an evaluation of $h_L$ for a line $L$ and evaluation point of its choice, we will instead allow the decoder to make ``line queries'' where the decoder reads the entire evaluation table of a function $h_L$ with one ``line query $L$''. Clearly, any $q$-query decoder in the standard query model can be simulated in the line query model with $\leq q$ queries, as any query to a specific evaluation of a local function $h_L$ can simply be replaced by a query to the entire line $L$. Notice that since the decoder makes line queries and each local function $\Had(f(S_L))$ is either completely replaced with $\bot$ or is untouched, we may further assume that the decoder receives $f \vert_L$ when it makes the line query $L$, as opposed to $\Had(f(S_L))$, as it can simply compute $\Had(f(S_L))$ from $f \vert_L$. Thus, we will view the decoder as having ``line query'' access to the polynomial $f$, where for every line it either receives $f \vert_L$ (the entire line) or $\bot$ (if the line has been erased). For a set of \emph{erased} lines $\cL$, we will use the notation $\Dec^{f, \cL}$ to indicate that the decoder has line query access to the function $f$ on all lines \emph{except} those in $\cL$. Namely, if the decoder queries a line $L$ and $L \in \cL$, then it receives $\bot$, and otherwise it receives $f \vert_L$. 

\parhead{Lower bound against ``line query'' decoders in the erasure model.} We will now show the following. For any ``line query'' decoder $\Dec$ making at most $d$ queries and for any $\alpha^* \in \F_{2^t}$, $x^* \in \F_{2^t}^n$, and $i^* \in [t]$, there is a set of $\delta$-fraction of lines $\cL$ that can be erased such that $\E_f[\Pr[\Dec^{f, \cL}(\alpha^*, x^*, i^*) = \pi(\alpha^* f(x^*))_{i^*}]] = \frac{1}{2}$, where the outer expectation is over the random choice of the polynomial $f$ and the inner probability is over the randomness of the decoder. In particular, this implies that there exists a polynomial $f$ of degree $\leq d$ such that $\Pr[\Dec^{f, \cL}(\alpha^*, x^*, i^*) = \pi(\alpha^* f(x^*))_{i^*}]] \leq \frac{1}{2}$, which will finish the proof.

Let us now argue the above claim. First, the set $\cL$ of lines will be all lines $L$ with $x^* \in L$, which we denote by $\cL_{x^*}$. We observe that the number of such lines is $(2^{tn} - 1)/(2^t - 1)$, which is a $\delta = 2^{-(t-1)n}$-fraction of all lines.
As we permit the decoder to be adaptive, the queries made by the decoder can be described as a depth $d+1$ decision tree (with $d$ layers of internal ``decision'' or ``query'' nodes and one layer of leaves or output nodes), where each decision node has a line $L$ that will be queried as well as a child for each possible evaluation table $f \vert_L$, and the output of the decoder is given by the leaves. Note that we may assume without loss of generality that the decoder makes exactly $d$ queries, so that the tree is a complete tree with depth exactly $d+1$. Furthermore, the decision tree is determined by the input $(\alpha^*, x^*, i^*)$ to the decoder and the choice $r$ of the decoder randomness. That is, for each input $(\alpha^*, x^*, i^*)$  and randomness $r$, there is a tree $T_{(\alpha^*, x^*, i^*,r)}$ that determines the queries that can be made when the decoder randomness is $r$. We use the notation $T_{(\alpha^*, x^*, i^*,r)}(f, \cL)$ to denote the output of the tree $T_{(\alpha^*, x^*, i^*,r)}$ when given access to $f$ on all lines not in $\cL$, which we note is equal to $\Dec^{f, \cL}(\alpha^*, x^*, i^*; r)$, the output of the decoder when the randomness is $r$. We thus have
\begin{flalign*}
&\E_{f}[\Pr[\Dec^{f, \cL_{x^*}}(\alpha^*, x^*, i^*) = \pi(\alpha^* f(x^*))_{i^*}]] = \E_f[\E_r[\1(T_{(\alpha^*, x^*, i^*,r)}(f, \cL) = \pi(\alpha^* f(x^*))_{i^*})]] \\
&=\E_r[\E_f[\1(T_{(\alpha^*, x^*, i^*,r)}(f, \cL) = \pi(\alpha^* f(x^*))_{i^*})]] \mcom
\end{flalign*}
using linearity of expectation, where the expectation over $f$ is over a random degree $\leq d$ polynomial and the expectation over $r$ is over the randomness $r$ of the decoder.

We will now argue that for any decision tree $T$ of depth $d+1$ that does not query lines in $\cL_{x^*}$, it holds that $\E_f[\1(T(f) = \pi(\alpha^* f(x^*))_{i^*})] = \frac{1}{2}$. (Note that we can replace $T_{(\alpha^*, x^*, i^*,r)}$ with an equivalent decision tree that never queries lines in $\cL_{x^*}$.) This follows from the following observation, which we will prove at the end of the subsection.

\begin{claim}
\label{obs:separatingpoly}
Let $x \in \F_{2^t}^n$ and let $L_1, \dots, L_q$ be lines in $\F_{2^t}^n$ that do not contain $x$. Then, there is a degree-$q$ polynomial $g \colon \F_{2^t}^n \to \F_{2^t}$ such that $g(y) = 0$ for all $y \in \cup_{i =1}^q L_i$ and $g(x) = 1$.
\end{claim}
For a set of $d$ lines $L_1, \dots, L_d$, let $g_{x^*, L_1, \dots, L_d}$ be the polynomial in \cref{obs:separatingpoly} (which may not be unique, but we choose an arbitrary fixed one for each choice of lines). Now, consider the following distribution: (1) sample $f$ uniformly at random, (2) let $L^{(f)}_1, \dots, L^{(f)}_d$ be the lines queried by $T(f)$ (which are not in $\cL_{x^*}$), and (3) output $f + \beta g_{x^*, L^{(f)}_1, \dots, L^{(f)}_d}$ where  $\beta \gets \F_{2^t}$ uniformly at random. Note that this is well-defined since $x \notin L^{(f)}_i$ for all $i \in [d]$, as $L^{(f)}_i \notin \cL_{x^*}$. We claim that this distribution is simply uniform on degree $\leq d$ polynomials, i.e., it is the same as choosing $f$ uniformly at random. To see this, let $S_f = \{f + \beta g_{x^*, L^{(f)}_1, \dots, L^{(f)}_d}\}_{\beta \in \F_{2^t}}$. Observe that for any $h \in S_f$, we have that $S_{h} = S_f$. Indeed, this follows because $g_{x^*, L^{(f)}_1, \dots, L^{(f)}_d}(y) = 0$ for all $y \in \cup_{i = 1}^d L^{(f)}_i$, and so it follows that $h$ agrees with $f$ on all lines $L^{(f)}_1, \dots, L^{(f)}_d$, and hence the decision tree must follow the same root-to-leaf path for both $h$ and $f$. In particular, this also implies that $T(f) = T(h)$ for all $h \in S_f$. It thus follows that the $S_f$'s partition the set of degree $\leq d$ polynomials into sets of size $2^t$. Hence, the above distribution simply first chooses a set $S_f$ in the partition uniformly at random, and then chooses a random polynomial in $S_f$, which is the same as choosing a uniformly random degree $\leq d$ polynomial.

Now, to finish the argument, we have that
\begin{flalign*}
&\E_f[\1(T(f) = \pi(\alpha^* f(x^*))_{i^*})] = \E_{f}[\E_{h \in S_f}[\E_{\beta}[\1(T(h) =  \pi(\alpha^* h(x^*))_{i^*})]]] \\
&=  \E_{f}[\E_{h \in S_f}[\E_{\beta}[\1(T(f) =  \pi(\alpha^* h(x^*))_{i^*})]]] = \frac{1}{2} \mcom
\end{flalign*}
where the third equality uses that $T(f) = T(h)$ for all $h \in S_f$ and the fourth inequality uses that $g_{x^*, L^{(f)}_1, \dots, L^{(f)}_d}(x^*) = 1$, so that $\pi(\alpha^* h(x^*))_{i^*})$ is distributed as a uniformly random bit when $h \gets S_f$.

Because the above argument holds for any decision tree $T$ that does not query lines in $\cL_{x^*}$, it follows that 
\begin{flalign*}
&\E_f[\Pr[\Dec^{f, \cL_{x^*}}(\alpha^*, x^*, i^*) = \pi(\alpha^* f(x^*))_{i^*}]] =\E_r[\E_f[\1(T_{(\alpha^*, x^*, i^*,r)}(f, \cL_{x^*}) = \pi(\alpha^* f(x^*))_{i^*})]] = \frac{1}{2} \mcom
\end{flalign*}
as $T(\cdot, \cL_{x^*})$ is a decision tree that cannot query lines in $\cL_{x^*}$. This shows that the decoder's success probability cannot exceed $1/2$ in expectation over a random codeword, and hence there exists a codeword where the success probability is $\leq 1/2$. This finishes the proof up to the proof of \cref{obs:separatingpoly}, which we do below.
\end{proof}

\begin{proof}[Proof of \cref{obs:separatingpoly}]
We will first show that for any line $L$ and any point $x \notin L$, there is a degree-$1$ polynomial $g_{x, L}$ satisfying $g_{x,L}(x) = 1$ and $g_{x,L}(y) = 0$ for all $y \in L$. Let $L = \{a + \lambda b\}$ where $a \in \F_{2^t}^n$, $b \in \F_{2^t}^n \setminus \{0^n\}$, and $\lambda \in \F_{2^t}$.
Since $x \notin L$, there exists $v \in \F_{2^t}^n$ such that $\ip{x - a, v} \ne 0$ and $\ip{b, v} = 0$. Indeed, this follows because $x \notin L$ implies that $x - a$ and $b$ are linearly independent, in which case the subspaces $\vspan\{b\}$ and $\vspan\{x - a, b\}$ are distinct subspaces of dimension $1$ and $2$, respectively. The vector $v$ is simply any vector in $(\vspan\{b\})^{\perp} \setminus (\vspan\{x - a, b\})^{\perp}$, which is nonempty since the two subspaces are distinct.

With the vector $v$, we now let $g_{x,L}(z) \defeq \ip{z - a,v}/\ip{x - a,v}$, which is a degree-$1$ polynomial that is well-defined since $\ip{x - a,v} \ne 0$. We have that $g_{x,L}(x) = 1$, and for any $y \in L$, i.e., $y = a + \lambda b$, we have that $g_{x,L}(a + \lambda b) = \lambda \ip{b, v}/\ip{x - a,v} = 0$ since $\ip{b,v} = 0$.

Finally, to finish the proof, we simply take $g_{x, L_1, \dots, L_q}$ to be $\prod_{i = 1}^q g_{x, L_i}$.
\end{proof}

\section{Lower Bound for Linear 2-RLDCs Over Any Finite Field}
\label{app:2rldc-lower-bound}
In this section, we extend \cref{mthm:rldcsoundnessthreshold} to any finite field $\F$, for the specific case of $q=2$. 
\begin{theorem}
  \label{thm:2rldc-lower-bound}
  Let $\Code\colon \F^k \to \F^n$
  be a linear $(2, \delta, 1, s)$-RLDC
  with a nonadaptive decoder.
  Then,
  there exists a linear $(2, \delta/2, 1, s)$-LDC
  $\Code' \colon \F^{k'} \to \F^n$
  where $k' \ge k - \floor{2/\delta}$
  (and analogously for 2-RLCCs implying 2-LCCs).
  By \cite[Theorem 1.4]{GoldreichKST06}, this implies that
  $n \ge 2^{\Omega_{\delta, s}(k) - \log_2(\abs{\F})}$.
\end{theorem}
\cref{thm:2rldc-lower-bound} thus extends the exponential lower bound of \cite{BlockBCGLZZ23} for $2$-query binary RLDCs to arbitrary finite fields.
It also extends the proof ideas to get the corresponding result that 2-RLCCs give 2-LCCs.
With \cref{thm:stronger-goldberg},
we can extend this lower bound to 2-RLDCs and 2-RLCCs with imperfect completeness and adaptive decoders.
\begin{corollary}
    Let $\Code\colon \F^k \to \F^n$
    be a linear $(2,\delta,1-\eps,s)$-RLDC.
    Then,
    there exists
    a linear $(2, \delta, 1, s + (2 + 1/(\abs{\F}-1))\eps)$-LDC
    $\Code' \colon \F^{k'} \to \F^n$
  where $k' \ge k - \floor{2/\delta}$
  (and analogously for 2-RLCCs implying 2-LCCs).
  By \cite[Theorem 1.4]{GoldreichKST06}, this implies that
  $n \ge 2^{\Omega_{\delta, s, \eps, \abs{\F}}(k) - \log_2(\abs{\F})}$.
\end{corollary}

To prove \cref{thm:2rldc-lower-bound}, we adapt the approach of \cite{BlockBCGLZZ23} to show that
every linear 2-RLDC (or 2-RLCC) over \emph{any} finite field
must also be a 2-LDC (or 2-LCC).\footnote{In fact, assuming linear structure
  eliminates much of the casework necessary in
  \cite{BlockBCGLZZ23}. It would be interesting to extend this proof
  to nonlinear codes over large alphabets.}
We will show that the relaxed local decoder has two {``modes,''}
one of which behaves like a non-relaxed local decoder,
and the other which necessarily has poor soundness.
Thus, if the 2-RLDC or 2-RLCC has decent soundness,
we can eliminate the latter case to
build a 2-LDC or 2-LCC.

At a high level, we can use the structure of a linear code
from \Cref{fact:canonical}
to argue that
every local view of the decoder
either looks like a Hadamard code (``smooth'' case)
or a repetition code (``nonsmooth'' case).
Then, we can observe that
the latter case must have high soundness error,
and so our 2-RLDC must essentially be a
Hadamard code, which is a 2-LDC.
\subsection{Smooth (Hadamard code)
  vs. nonsmooth (repetition code) cases}
Let $\Dec(\cdot)$ be the relaxed local decoder (or corrector) for $\Code$
satisfying the assumed decoding radius, completeness,
and soundness parameters.
Assuming that the decoder is nonadaptive,
we can assume that $\Dec^y(i)$ behaves as follows:
\begin{enumerate}
\item Sample a query pair $(j,\ell) \leftarrow \mathcal{Q}_i$
  with a corresponding decoding function
  $f_{j,\ell}^i\colon \F^2 \to \F \cup \{\bot\}$.
\item Query $y$ at indices $j,\ell$ and return
  the evaluation of $f_{j,\ell}^i(y_j, y_\ell)$.
\end{enumerate}
Consider each query pair $(j,\ell)$ from the support of $\mathcal{Q}_i$.

\begin{itemize}
    \item Suppose $\Code$ is an RLDC.
    Then
    we say that $j$ is \emph{fixed} by $i$
if there is some $\alpha \in \F$
such that $\Code(\msg)_j = \alpha \msg_i$ for all $\msg$.
\item If $\Code$ is an RLCC,
    then
    we say that $j$ is \emph{fixed} by $i$
if there is some $\alpha \in \F$
such that $\Code(\msg)_j = \alpha \Code(\msg)_i$ for all $\msg$.
\end{itemize}

Let $S_i \subseteq [n]$ be the set of codeword indices
fixed by $i$.
Note that because the encoding is linear,
every $j \in [n]$ is either unfixed,
or fixed by exactly one index.
Then, we can consider each case:
\begin{enumerate}
\item If both $j$ and $\ell$
  are not fixed by $i$,
  then the decoding function $f_{j,\ell}^i$ cannot return $\bot$,
  because $\Code\rvert_{\{j,\ell\}}$ has dimension $2$
  and any local view can be completed to a valid codeword
  due to \cref{fact:restriction-support}.
  Thus, $f_{j,\ell}^i$ always returns a linear combination of
  its two inputs, which behaves like the local decoder
  of a \emph{Hadamard code}.
  This is the {``smooth''} case
  described in \cref{sec:proofstrategy}.

\item If both $j$ and $\ell$ are fixed by $i$, then
  $\Code\rvert_{\{j,\ell\}}$ has dimension $1$.
  The truth table of the
  decoding function $f_{j,\ell}^i$ has
  exactly $\abs{\F}^2 - \abs{\F}$
  entries that are $\bot$.
  Because both entries are multiples of the $i$-th message symbol,
  this local view looks like a \emph{repetition code}.
  This is necessarily the {``nonsmooth''} case
  because these codeword indices only contain information on a single
  message index.

\item If $j$ is fixed by $i$ but $\ell$ is not fixed by $i$
  (or vice versa),
  then $\Code\rvert_{\{j,\ell\}}$ has dimension $2$
  and so $f_{j,\ell}^i$ never returns $\bot$.
  In addition, by perfect completeness,
  the decoding function must take the form
  $f_{j,\ell}^i(y_j, y_\ell) = \alpha^{-1} y_j$,
  and so it does not depend on the $\ell$-th symbol at all.
  Hence, we can treat $(j,\ell)$ as if it queried only $j$, which is a repetition code case.
\end{enumerate}

Thus, without loss of generality, either both queries are fixed by $i$
(and the decoder essentially tests equality between two copies of $i$)
or both are not fixed by $i$ (and the decoder cannot test and must always return a linear combination of the two symbols).
We can show that as long as $\abs{S_i}$ is not too large,
the repetition code case can be fooled with certainty
(compare to the higher query case in \cref{lem:fooling}, where the tests can be more sophisticated).
This implies that the
decoder has good soundness
when conditioned on choosing a Hadamard-like query case, where it never returns $\bot$.
\begin{claim}
  \label{claim:2rldc-hadamard}
  Suppose $\Code$ is a 2-RLDC satisfying 
  the premise of \cref{thm:2rldc-lower-bound}
  and suppose $i \in [k]$ such that $\abs{S_i} \leq \delta n / 2$.
  Then, for all $y$ such that $\Delta(y, C(\msg)) \leq \delta n / 2$,
  \begin{equation*}
    \Pr[\Dec^{y}(i) = \msg_i \mid \text{Hadamard-like query case}]
    \geq 1 - s\mper
  \end{equation*}
  Analogously for a 2-RLCC $\Code$ 
  and for $i \in [n]$
  satisfying the same conditions,
  \begin{equation*}
    \Pr[\Dec^{y}(i) = \Code(\msg)_i \mid \text{Hadamard-like query case}]
    \geq 1 - s\mper
  \end{equation*}
\end{claim}
\begin{proof}
  Let $E\coloneq j \in S_i \land k \in S_i$
  be the event of the {``repetition code''} case.
  We will bound the behavior of the decoder
  conditioned on $\lnot E$ (i.e., conditioned on the Hadamard case)
  by tampering with $y$ and introducing errors
  designed to completely fool the repetition code case.
  We mildly reduce the decoding radius from $\delta$
  to $\delta/2$ to give us the breathing room to introduce these errors,
  which are only used for the analysis.

  Begin by finding a codeword $c$
  that differs in the desired symbol:
  \begin{itemize}
      \item If $\Code$ is a 2-RLDC,
      then let $c \coloneqq \Code(\msg + e_i)$,
      so that it differs in the $i$-th message symbol.
      \item If $\Code$ is a 2-RLCC,
      let $c$ be any arbitrary codeword 
      such that $c_i \neq \Code(\msg)_i$;
      one must exist unless $\Code\rvert_{\{i\}} = \{0^n\}$.
  \end{itemize}
  Then, let $y'$ be the string formed by taking $y$
  and then setting $y'\rvert_{S_i} = c\rvert_{S_i}$,
  i.e., replace the symbols at $S_i$ with those from $c$.
  Using the triangle inequality, $\Delta(y', \Code(\msg)) \le \delta n$,
  and so $y'$ is subject to the soundness error property of $\Code$:
  \begin{itemize}
      \item If $\Code$ is a 2-RLDC, then
        \begin{align*}
    1 - s \le 
     \Pr[E]
      \cdot \Pr[\Dec^{y'}(i) \in \{\msg_i, \bot\} \mid E]
      + \Pr[\lnot E]
      \cdot \Pr[\Dec^{y'}(i) \in \{\msg_i, \bot\} \mid \lnot E] \mper
  \end{align*}

  \item If $\Code$ is a 2-RLCC, then
        \begin{align*}
    1 - s \le 
     \Pr[E]
      \cdot \Pr[\Dec^{y'}(i) \in \{\Code(\msg)_i, \bot\} \mid E]
      + \Pr[\lnot E]
      \cdot \Pr[\Dec^{y'}(i) \in \{\Code(\msg)_i, \bot\} \mid \lnot E] \mper
  \end{align*}
  \end{itemize}

  However, conditioned on event $E$,
  our decoder $\Dec$ makes two queries to indices in $S_i$,
  and hence sees a local view indistinguishable
  from $c$, which we purposely chose to trick our decoder.
  By the perfect completeness property,
  the decoder must return $\msg_i + 1 \neq \msg_i$ (or $c_i \neq \Code(\msg)_i$) with probability $1$.
  This gives a lower bound on the conditional probability of
  the decoder returning the right answer:
  \begin{equation*}
    1 - s \leq
    \Pr[\lnot E]
    \cdot \Pr[\Dec^{y'}(i) \in \{\msg_i, \bot\} \mid \lnot E]
    \leq \Pr[\Dec^{y}(i) = \msg_i \mid \lnot E]\mcom
  \end{equation*}
  or for a 2-RLCC, $\Pr[\Dec^{y}(i) = \Code(\msg)_i \mid \lnot E] \ge 1 - s$.
  We use two key observations here:
  \begin{enumerate}
  \item Recall that $y\rvert_{[n] \setminus S_i} = y'\rvert_{[n] \setminus S_i}$.
    Conditioned on on $\lnot E$,
    both queries are outside of $S_i$, and since $y' \vert_{S_i} = y \vert_{S_i}$, we can replace $y'$ with $y$ on the right hand side.
  \item In addition, the decoder cannot return $\bot$
    because every possible pair of queried values
    will agree with some codeword.
    Hence, $\Dec(\cdot)$ must return $\msg_i$ (or $\Code(\msg)_i$) to satisfy perfect completeness.
    \qedhere
  \end{enumerate}
\end{proof}
\subsection{Repetition case is rare}
If $\abs{S_i} \leq \delta n / 2$
for every $i \in [k]$
(or every $i \in [n]$),
then \cref{claim:2rldc-hadamard}
would prove that $\Code$ is a $(2, \delta/2, 1, s)$-LDC (or LCC),
because we can simply condition our decoder's query distribution
to pick
a Hadamard-like local view
and get the required soundness error property.
This may not be true for every $i$,
but we can use a simple counting argument
to show that it holds for
nearly all of the message indices $i$.
Thus, $\Code$
itself is not necessarily a 2-LDC (or 2-LCC),
but it does {``contain''} a 2-LDC (or 2-LCC) with nearly the same parameters.
\begin{claim}
  Let $X = \{i : \abs{S_i} > \delta n / 2 \}$.
  Then, $\abs{X} \leq 2/\delta$.
\end{claim}
\begin{proof}
  Recall that a codeword index $j$ cannot be fixed by two different (message for RLDC or codeword for RLCC) indices.
  Thus, the sets $S_i$ are mutually disjoint, so
  \begin{equation*}
    \delta \abs{X} n / 2 < \sum_{i \in X} \abs{S_i} \le n
    \implies \abs{X} < 2 / \delta\mper
  \end{equation*}
  Note that the sets $S_i$ may not be mutually disjoint
  if the code is nonlinear.
  Indeed, in this case a codeword index could be fixed
  by many different message indices,
  which (for the binary case) requires a careful analysis
  in \cite{BlockBCGLZZ23}.
\end{proof}

Now, we have the tools to prove \cref{thm:2rldc-lower-bound}.
We will remove the indices in $X$
from the code, either by fixing them to zero
for the RLDC case
or by adding linear constraints setting them to zero for the RLCC case.
Then, every remaining message or codeword index
either satisfies \cref{claim:2rldc-hadamard}
or has its value fixed.
\begin{proof}[Proof of \cref{thm:2rldc-lower-bound}]
Start by defining $k' \coloneq k - \floor{2/\delta}$.
For $\Code$ which is an RLDC,
without loss of generality,
  $X = \{k' + 1, k' + 2, \dots, k\}$.
  Let $\Code'\colon \F^{k'} \to \F^n$
  be defined as $\Code'(\msg) = \Code((\msg, 0^{k-k'}))$,
  i.e., append $k - k' = \abs{X}$ zeroes to the message
  and encode using $\Code$.
  Then, every $i \in [k']$
  satisfies $\abs{S_i} \leq \delta n / 2$,
  and so \cref{claim:2rldc-hadamard}
  gives us a local decoder for all $i \in [k']$
  which never returns $\bot$.
  This gives the required $(\delta/2, s)$-soundness error property.
  The query complexity and perfect completeness properties
  are inherited from the original decoder,
  showing that $\Code'$
  is a $(2, \delta/2, 1, s)$-LDC.
  At last, apply the 2-LDC lower bound
  of~\cite[Theorem~1.4]{GoldreichKST06}
  which gives $n \ge 2^{(1/2 - s)\delta k' / 16 - 1 - \log_2 {\abs{\F}}}$.

  For $\Code$ which is an RLCC, 
  take an arbitrary parity check matrix $B \in \F^{(n-k) \times n}$
  for $\Code$,
  and then add each $e_i \in \F^n$ for $i \in X$ as rows to $B$.
  This defines a new code $\Code' \subseteq \Code$
  with dimension at least $k'$ where the indices at $X$ are always zero,
  and we can pick an arbitrary basis to get the encoding map $\Code' \colon \F^{k'} \to \F^n$.
  If $i \in [n]$ is not in $X$, then it satisfies \cref{claim:2rldc-hadamard}
  yielding a local corrector for $i$ that never returns $\bot$;
  if $i \in [n]$ is in $X$, we can now use a trivial local corrector that always returns zero.
  Thus, $\Code'$ is a $(2, \delta/2, 1, s)$-LCC.  
\end{proof}

\section*{Acknowledgements}
We thank Tselil Schramm, Guy Bresler, Sam Hopkins, and Luca Trevisan for organizing a
wonderful workshop at the Banff International Research Station that led to the start of this project. We also thank the anonymous reviewers for their insightful comments that have improved the presentation of the
paper.

This material is based upon work supported by the National Science Foundation
under Grant No.\ DMS-2424441 and the National Science Foundation Graduate
Research Fellowship Program under Grant No.\ DGE-2137420. Any opinions, findings,
and conclusions or recommendations expressed in this material are those of the authors
and do not necessarily reflect the views of the National Science Foundation.

\bibliographystyle{alpha}
\bibliography{references.bib}

\appendix

\section{Linear LDCs and LCCs Have Strong Soundness}
\label{app:strongsoundness}
In this appendix, we show that any linear LDC or LCC satisfying \cref{def:ldc,def:lcc} additionally has strong soundness (\cref{def:strongldc}). More formally, we will show the following lemma.
\begin{lemma}
Let $\Code \colon \F^k \to \F^n$ be a linear $(q, \delta, c, s)$-LDC (LCC) with $s < 1 - \frac{1}{\abs{\F}}$. Then, there is a decoder $\Dec(\cdot)$ satisfying the following conditions:
\begin{enumerate}[(1)]
\item ($q$-queries) for any $y$ and $i$, the algorithm $\Dec^{y}(i)$ reads at most $q$ indices of $y$
\item (perfect completeness) for all $b \in \F^k$ and $i \in [k]$, $\Pr[\Dec^{\Code(b)}(i) = b_i] = 1$,
\item ($(\delta/q)$-strong soundness) for all $b \in \F^k$, $i \in [k]$, and all $y \in \F^n$ with $\Delta(y, \Code(b)) \leq \delta n/q$, $\Pr[\Dec^{y}(i) \ne b_i] \leq \frac{q\Delta(y, \Code(b))}{\delta n}$.
\end{enumerate}
\end{lemma}
\begin{proof}
By~\cite{Dvir16a}, given any $(q, \delta, c, s)$-LDC, there exist collections $\cH_1, \dots, \cH_k$ where (1) each $\cH_i$ is a set of $\abs{\cH_i} \geq \delta n/q$ $q$-sparse vectors in $\F^n$ with disjoint support, i.e., any $v, v' \in \cH_i$, $\supp(v) \cap \supp(v') = \emptyset$, and (2) for every $b \in \F^k$, $i \in [k]$, and $v \in \cH_i$, it holds that $b_i = \sum_{j = 1}^n v_j x_j$ where $x = \Code(b)$. Note that the latter sum is actually over at most $q$ indices, as $\abs{\supp(v)} \leq q$.

Consider the decoder $\Dec^{y}(i)$ that behaves as follows: on input $i$, sample $v \gets \cH_i$ uniformly at random, read $y_j$ for each $j \in \supp(v)$, and output $\sum_{j \in \supp(v)} v_j y_j$. If $y_j = x_j$ for all $j \in \supp(v)$, where $x = \Code(b)$, then by the above, the decoder outputs $b_i$.

Finally, $b \in \F^k$, $x = \Code(b)$, and let $y \in \F^n$ with $\Delta(y, \Code(b)) \leq \delta n/q$. Let $S$ denote the set of coordinates $j \in [n]$ where $y_j \ne \Code(b)_j$. Because $\supp(v)$'s are disjoint for $v \in \cH_i$, it follows that 
\begin{equation*}
\Pr_{v \gets \cH_i}[\supp(v) \cap S \ne \emptyset] \leq \E_{v \gets \cH_i}[\abs{\supp(v) \cap S}] = \frac{1}{\abs{\cH_i}}\sum_{v \in \cH_i} \abs{\supp(v) \cap S} = \frac{\abs{(\cup_{v \in \cH_i} \supp(v)) \cap S}}{\abs{\cH_i}} \leq \frac{\abs{S}}{\abs{\cH_i}} \mper
\end{equation*}
As $\abs{\cH_i} \geq \delta n/q$, it follows that the probability the decoder outputs $b_i$ is at least $1 - \frac{q \abs{S}}{\delta n}$, as required.

To prove the statement for LCCs, we use~\cite{Dvir16b} instead of~\cite{Dvir16a}, and the rest of the proof follows immediately.
\end{proof}

\end{document}